RESEARCH ARTICLE

# One-step decellularization of porcine uterine tissue for developing alginate–decellularized uterine ECM hydrogel for uterine tissue engineering

**Abbas Fazel Anvari-Yazdi[1]*, Kobra Tahermanesh[2], Maryam Ejlali[3], Louison Blivet-Bailly[1,4], Vatsala Singh[1,5], Bishnu Acharya[6], Daniel J. MacPhee[7], Ildiko Badea[2]*, and Xiongbiao Chen[1,8]***

[1]Division of Biomedical Engineering, University of Saskatchewan, 57 Campus Dr, Saskatoon, Canada

[2]Department of Obstetrics and Gynecology, School of Medicine, Iran University of Medical Sciences (IUMS), Tehran, Iran

[3]College of Pharmacy and Nutrition, University of Saskatchewan, Saskatoon, Canada

[4]École Supérieure de Physique et de Chimie Industrielles de la Ville de Paris (ESPCI Paris), PSL Research University, 10 rue Vauquelin, Paris, France

[5]Department of Biomedical Engineering, University of Michigan, Ann Arbor, Michigan, United States of America

[6]Department of Chemical and Biological Engineering, University of Saskatchewan, 57 Campus Drive, Saskatoon, Canada

[7]Department of Veterinary Biomedical Sciences, Western College of Veterinary Medicine, University of Saskatchewan, 52 Campus Drive, Saskatoon, Canada

[8]Department of Mechanical Engineering, University of Saskatchewan, 57 Campus Dr, Saskatoon, Canada

***Corresponding authors:**
Abbas Fazel Anvari-Yazdi
(fazel.a@usask.ca)
Ildiko Badea
(ildiko.badea@usask.ca)
Xiongbiao Chen
(xbc719@usask.ca)

## Abstract

Decellularized uterine extracellular matrix (dUECM) is promising for uterine tissue engineering because of its inherent bioactivity and structural complexity. However, transforming dUECM into porous, functional 3D constructs remains a significant challenging. This study aimed to: (1) synthesize dUECM using a modified decellularization protocol and formulate it into a hydrogel ink, and (2) to fabricate 3D-printed constructs from this ink to assess their capacity to support human uterine myometrial cell growth *in vitro*. Porcine uterine tissues were decellularized using 1% Triton™ X-100 with varying concentrations of sodium dodecyl sulfate (SDS) (0.1–1.5%) for 48–72 h. The resulting dUECM was characterized using DNA and glycosaminoglycan (GAG) quantification, Picrosirius Red-polarized light microscopy, routine histology, scanning electron microscopy (SEM), Fourier-transform infrared spectroscopy (FTIR), Raman spectroscopy, and thermogravimetric analysis (TGA). To prepare the ink, dUECM powder was enzymatically digested with pepsin and subsequently blended with 2% and 3% alginate to obtain a printable hydrogel formulation. Constructs were fabricated using extrusion-based 3D printing and assessed for filament fidelity, swelling, degradation behavior, and mechanical properties. Biocompatibility was evaluated using hTERT-HM myometrial cells through MTT metabolic assays, Live/Dead staining, and immunohistochemical α-SMA staining. The optimal protocol (1%

Triton™ X-100 + 1% SDS for 48 h) reduced DNA to 51.3 ± 9 ng/mg while retaining a high level of GAGs (54.9 ± 7.6 μg/mg). Preservation of the ECM structure was confirmed by spectroscopy analyses. The 3% Alg + 1.5% dUECM hydrogel exhibited suitable printability (1.5 ± 0.2), swelling capacity (47 ± 12%), degradation resistance (94 ± 18% mass retention), and mechanical strength (decreasing from 323 kPa to 175 kPa over 14 days), along with high viability and proliferation (258 ± 13%). The developed dUECM-based hydrogel supports 3D bioprinting with strong mechanical and biological performance, offering a promising platform for uterine tissue engineering.

***Keywords:*** Tissue engineering; Hybrid hydrogel; Scaffolds; 3D-extrusion printing; Uterine smooth muscle; Extracellular matrix

## 1. Introduction

Uterine tissue engineering has the potential to transform the treatment of infertility, congenital anomalies, uterine dysfunction, and pregnancy complications. Cesarean section (CS) is one of the most frequently performed surgical procedures worldwide, accounting for approximately 21.1% of all births.[1] While CS is life-saving, it can lead to complications like uterine scarring, impaired healing uterine niches, and subsequent fertility problems.[2-5] To address these CS-related complications and other forms of uterine dysfunction, uterine tissue engineering using biomaterial-based scaffolds or constructs has recently emerged as a promising therapeutic strategy.[6]

The development of functional constructs for uterine tissue engineering has gained considerable attention, but several challenges persist. A major challenge is developing biomaterials with suitable biocompatibility, mechanical integrity, and reproducibility. Various biomaterials have been investigated, and among them, decellularized uterine extracellular matrix (dUECM)—the native uterine tissue rendered acellular while retaining its ECM components—has emerged as a particularly promising option due to its essential structural and biochemical cues.[7-9] Recent work has further demonstrated that uterine-derived dECM can mediate endometrial regeneration and enhance fertility outcomes, and that uterus-mimetic ECM microenvironments represent a promising strategy for engineering female reproductive tissues.[10,11] The native extracellular matrix (ECM) consists of hundreds of proteins, glycosaminoglycans (GAGs), proteoglycans, growth factors, and cytokines, collectively referred to as the "matrisome."[12] These components regulate key cellular behaviors, including adhesion, proliferation, migration, and differentiation. Decellularization of uterine tissue aims to remove immunogenic cellular components while preserving bioactive ECM molecules, offering a unique advantage over synthetic or non-tissue-specific natural biomaterials.[8,13,14] Additionally, native tissue-derived dUECM requires no exogenous crosslinking agents, thereby minimizing and eliminating risks associated with chemical or irradiation-based stabilization methods.[15]

Decellularization protocols play a critical role in determining the quality of the resulting dUECM by balancing effective cell removal with preservation of ECM structure and biochemical integrity.[16] A wide range of physical, chemical, and enzymatic techniques have been developed, often in combination, to achieve optimal decellularization.[17] Among chemical agents, detergents such as octyl-phenoxy-polyethoxy-ethanol (Triton™ X-100) and sodium dodecyl sulfate (SDS) have long been recognized as effective decellularizing agents.[18] Triton™ X-100, a non-ionic detergent, preserves ECM morphology and bioactivity but is limited in its ability to fully remove cellular material,[19] whereas SDS, an ionic detergent, removes cellular debris and DNA more effectively but may denature proteins and compromise ECM biochemical composition if used for prolonged periods.[20] Previous studies have identified 1% Triton™ X-100 and 1% SDS as commonly used concentrations; however, the effects of SDS concentrations below or above 1% remain insufficiently explored.[21-28] Notably, these detergents are often used in separate sequential steps, with each step taking 24–48 h, and total decellularization processes frequently spanning up to a week. The prolonged exposure to water, detergents, and mechanical agitation significantly alters ECM biochemical composition, leading to the loss of crucial proteins such as glycosaminoglycans.[29] To address these challenges, this study aimed to develop and evaluate a simplified, one-step decellularization protocol combining 1% Triton™ X-100 with variable SDS concentrations to reduce processing time while preserving essential ECM components.

Despite its bioactivity, dUECM exhibits poor mechanical strength, limiting its suitability for fabrication into porous three-dimensional (3D) constructs. Extrusion-based 3D printing has emerged as a powerful technique to fabricate spatially controlled, porous constructs for tissue engineering.[8] In this context, alginate (Alg) has been widely used due to its suitable printability, biocompatibility, biodegradability, and ion-mediated crosslinking capability.[30-32] However, Alg lacks inherent cell-adhesion motifs, which limit its ability to support robust outcomes.[33] Therefore, combining dUECM with Alg to integrate bioactive cues with optimal printability represents a rational and synergistic strategy, yet this approach has not been reported in the literature. In uterine tissue engineering, uterine smooth muscle cells play an essential role in tissue remodeling, contractility, and structural integrity of the uterus. It has been demonstrated that incorporating smooth muscle cells into engineered constructs enhances the ability to replicate the physiological behavior of the uterus more effectively.[34]

This study presents the development of a bioactive hybrid hydrogel composed of dUECM and Alg for use in 3D bioprinting. We describe the one-step decellularization process, biochemical and structural characterization of the resulting dUECM, and the formulation of printable Alg:dUECM biomaterial ink. The extrusion printing process, as well as the hydrogel's mechanical properties, swelling behavior, degradation profile, and printability, are systematically evaluated. Additionally, *in vitro* biocompatibility was assessed by seeding hTERT-HM cells onto Alg:dUECM hydrogels, followed by analysis of cell viability and proliferation over time.[35] The following sections provide detailed descriptions of the experimental design and evaluation methods.

## 2. Materials and methods

### 2.1. Decellularization of porcine uterine tissue

Whole reproductive tissue was procured from ~1-year-old pigs weighing ~200 kg ($n$ = 8) at a provincially licensed slaughterhouse. Because the tissues were collected post-mortem as abattoir biowaste, no institutional ethical approval was required according to our institutional animal care guidelines. The tissues were immediately submerged in 1× sterile phosphate-buffered saline (PBS; Sigma-Aldrich, Cat.# P4417) and transported to the laboratory within 30 min to the lab for processing. The uterine tissues were isolated and stored at −80 °C freezer for subsequent use. For decellularization, the frozen tissues were transversely sliced into ~2 mm discs and thawed overnight at 4 °C, and then washed with sterile Dulbecco's phosphate-buffered saline (DPBS pH 7.4; Gibco, Cat.# 21600) with 1% antibiotic/antimycotic solution (Anti-Anti; Gibco, Cat.# 15240062). Each set of sliced tissues (30 g/500 mL DPBS) was transferred into 1,000 mL Erlenmeyer flasks, placed on an orbital shaker, and sealed with Parafilm® to prevent contamination. The tissues were continuously agitated at 200 rpm for 24 h at 4 °C to remove remaining blood residue. The DPBS solution was replaced every 8 h. Decellularization was performed using a combined detergent solution containing sodium dodecyl sulfate (SDS) at 0.1%, 0.5%, 1%, or 1.5% (Sigma-Aldrich, Cat.# L3771) and Triton™ X-100 at 1% (Fisher Bioreagents, Cat.# BP151). Both detergents were added simultaneously to ensure uniform exposure. Tissue samples were treated for either 48 h or 72 h, with detergent solutions refreshed every 8 h. Following detergent treatment, tissues were washed in ultrapure deionized water (Milli-Q) for 72 h at 4 °C to remove detergent residues. In total, eight decellularization conditions were evaluated. Figure 1 summarizes the workflow and characterization steps.

### 2.2. Quantification of residual SDS in decellularized tissues

Residual SDS in the decellularized tissues was quantified using the methylene blue active substances (MBAS) assay, adapted from previously established protocols.[36-38] Briefly, approximately 1 g of freeze-dried decellularized tissue was cryo-ground in liquid nitrogen and suspended in 10 mL of UltraPure™ water (Invitrogen™, Cat.# 10977015) and placed on a vertical rotary mixer for seven days at 4 °C. An aliquot of 2.5 mL of this suspension was mixed with 7.5 mL of ethanol to precipitate insoluble material, and the mixture was filtered. Then, 2 mL of the resulting filtrate was combined with 2 mL of methylene blue solution prepared from Methylene Blue hydrate (Sigma-Aldrich, Cat.# 66720; 25 mg/100 mL distilled water) and mixed thoroughly, followed by extraction with 4 mL chloroform (Sigma-Aldrich, Cat.# 1024451000) to isolate the methylene blue–SDS ion-pair complex. The mixture was then incubated overnight at room temperature in the dark to allow complete formation and stabilization of the ion-pair complex.

After phase separation, 1 mL of the chloroform layer was diluted to 10 mL with chloroform, and absorbance was measured at 650 nm using an Agilent 8,453 UV–Visible Spectrophotometer equipped with quartz cuvettes. SDS concentrations were determined from a calibration curve generated using serial dilutions of SDS (0–10 mg per 2.5 mL), processed in parallel using the same MBAS extraction procedure. Final SDS quantities were normalized to the dry weight of the extracted tissue and expressed as µg SDS per mg dry tissue. All measurements were performed in triplicate.

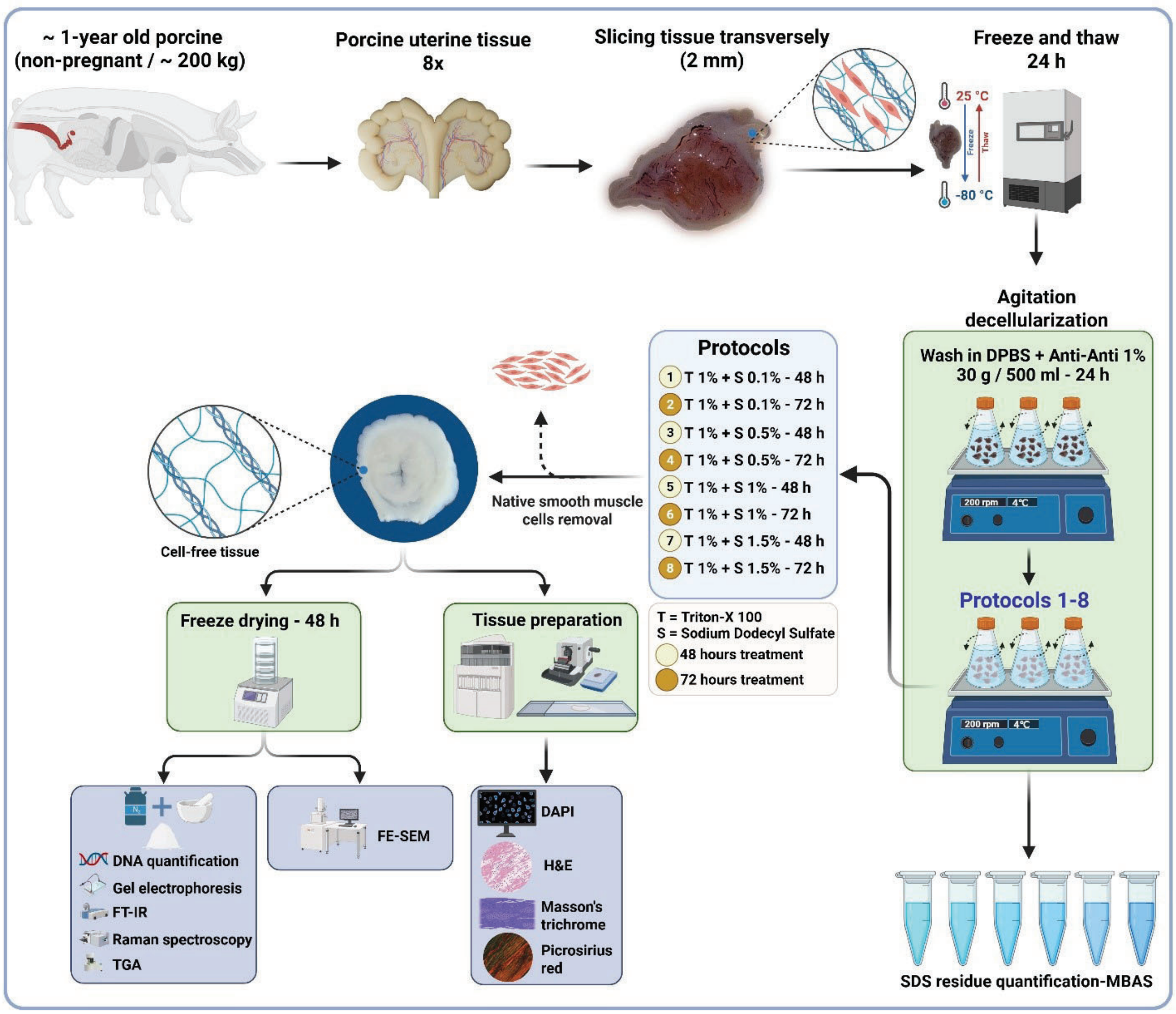

**Figure 1.** Porcine uterine decellularization protocols with a flow diagram of decellularization protocols and techniques used to quantify the efficiency of decellularization procedures

Abbreviations: Anti-Anti: Antibiotic/antimycotic solution; DAPI: 4′,6-diamidino-2-phenylindole; DPBS: Dulbecco's phosphate-buffered saline; FE-SEM: Field-emission scanning electron microscopy; FT-IR: Fourier-transform infrared spectroscopy; H&E: Hematoxylin and eosin; TGA: Thermogravimetric analysis.

The SDS residue (μg/mg dry tissue) was calculated using:

$$\text{SDS residue } (\mu\text{g/mg}) = \frac{DF \times V_{\text{chloro}} \times 10^{6}}{W_{\text{dry}}} \quad (1)$$

where

- $C_{SDS}$ = SDS concentration obtained from the calibration curve (mg/mL),
- $DF$ = 10 = dilution factor of the chloroform extract,
- $V_{\text{chloro}}$ = 4mL = volume of chloroform used for extraction,
- $W_{\text{dry}}$ = dry weight of tissue (mg)

The results were analyzed with one-way analysis of variance (ANOVA) with Tukey multiple comparisons test.

## 2.3. Microstructural analysis

### *2.3.1. DNA quantification and fragment-size determination*

Decellularized tissues were frozen and lyophilized to a constant mass. From the resulting dry material, 10 mg was weighed for each biological replicate ($n$ = 3 per protocol). DNA was extracted using the DNeasy® Blood and Tissue Kit (Qiagen, Cat.# 69504) according to the manufacturer's instructions. DNA concentration (ng DNA per mg dry tissue) was measured in technical triplicates using a NanoDrop® 2000c Spectrophotometer at 260 nm (Thermo

Scientific).

DNA fragment size was assessed by loading samples mixed with DNA Gel Loading Dye (Thermo Scientific, R0611) onto a 2% agarose gel containing SYBR Safe (Invitrogen, Cat.# S33102). Electrophoresis was performed for 90 min (60 min at 75 V, followed by 30 min at 90 V). A 1 kb Plus DNA Ladder (Invitrogen™, Cat.# 10787018) served as the molecular reference. Gels were imaged using a Bio-Rad ChemiDoc™ MP imaging system.

### 2.3.2. Glycosaminoglycans quantification

For glycosaminoglycans (GAGs) analysis, decellularized tissues were frozen and lyophilized to a constant mass. From the resulting dry material, 100 mg was weighed for each biological replicate ($n$ = 3 per protocol). Samples were digested in Proteinase K solution (20 µL/mg tissue; Qiagen, Cat.# 19131; pH 8) and incubated overnight at 56 °C, followed by enzyme inactivation at 90 °C for 10 min.

GAG content was quantified using the 1,9-dimethylmethylene blue (DMMB) assay (Sigma, Cat.# 341088) at pH 1.5. For each biological replicate, 20 µL of the digest was added to 200 µL of DMMB dye and dispensed into technical replicate wells ($n$ = 6) of a 96-well plate. Absorbance was read immediately at 650 nm using a SpectraMax 190 microplate reader (Molecular Devices). A standard curve was generated using heparin (Sigma-Aldrich, Cat.# H3393-250KU). GAG content was calculated and reported as µg GAG per mg dry tissue.

### 2.3.3. H&E and DAPI staining

Native and decellularized tissues (protocols 1–8) were fixed in 10% neutral buffered formalin (NBF; Sigma-Aldrich, Cat.# HT501128) for 24 h at ambient temperature, embedded in paraffin, and sectioned at 10 µm thickness. Cell removal was assessed using hematoxylin and eosin (H&E) and 4',6-diamidino-2-phenylindole (DAPI; Sigma, Cat.# MBD0015) staining.

For DAPI staining, sections were deparaffinized, rehydrated, incubated with DAPI stain, and mounted with ProLong™ Diamond Antifade Reagent (Invitrogen, Cat.# P36931). Slides were covered with micro cover glass (VWR, Cat.# 48404-454) and imaged using an automated microscope (BioTek® Lionheart LX). All images, including native tissue controls, were captured under identical exposure, gain, and illumination settings to ensure consistent nuclear detection.

### 2.3.4. Masson's trichrome staining

Masson's trichrome staining was performed to evaluate collagen distribution in native and decellularized tissues using a ready-to-use kit (Sigma-Aldrich, Cat.# HT15). Paraffin-embedded 10-µm sections were deparaffinized in xylene and rehydrated through graded ethanol. Sections were incubated in Bouin's solution (Sigma, Cat.# HT10132) at 56 °C for 15 min to enhance staining, rinsed in tap water, and stained with Weigert's iron hematoxylin for 5 min. After washing, slides were stained with Biebrich scarlet–acid fuchsin for 5 min, followed by aniline blue for 5 min. The reaction was stabilized with 1% acetic acid for 2 min. Slides were then rinsed, dehydrated, cleared in xylene, and cover-slipped. Under bright-field microscopy, collagen fibers appeared blue, cytoplasm red, and nuclei black.

### 2.3.5. Picrosirius Red staining

Collagen fiber organization and birefringence were assessed using Picrosirius Red staining. Native and decellularized tissues were paraffin-embedded and sectioned at 10 µm thickness. Staining was performed using the Abcam Picrosirius Red Stain Kit (Cat.# ab150681) following the manufacturer's instructions, with the addition of a hematoxylin pre-staining step to visualize possible nuclei.

Briefly, deparaffinized and rehydrated sections were first stained with hematoxylin for nuclear contrast, rinsed, and then incubated with Picrosirius Red solution to label collagen fibers. Slides were washed in acidified water, dehydrated, cleared, and cover-slipped. Stained sections were examined using a Leica DM2500 P Reflected & Transmitted Polarizing Light Microscope, enabling visualization of collagen birefringence and fiber alignment. Under polarized light, collagen fibers exhibited characteristic color patterns, with thicker fibers appearing red/orange, thinner fibers appearing green/yellow, and smooth muscle cells in gray.

### 2.3.6. Scanning electron microscopy

The microstructure of the decellularized tissues was investigated using field-emission scanning electron microscopy (FE-SEM; Hitachi SU8010). Samples were fixed using 2.5% glutaraldehyde (GA; Polysciences Inc., Cat.# 00376) at 4 °C overnight, then gradually dehydrated by an increasing concentrations of ethanol (10%, 30%, 60%, 90%, and 100%).[39,40] After dehydration, the samples were immersed into 1:2 and 2:1 hexamethyldisilizane (HMDS; Thermo Scientific, Cat.# A15139. AP): absolute ethanol, respectively, for 20 min and then 100% HMDS solution overnight. Samples were air-dried in a fume hood. For 3D-bioprinted constructs, the samples were frozen at −80 °C overnight and freeze-dried (LACONCO) at −50 °C under 0.05 mBar pressure for 48 h. Eventually, specimens were mounted on aluminum stubs with double-sided carbon tape and coated with 10 nm of gold (Quorum Q150TES Sputter Coater) as the conductive layer. For imaging, the voltage was fixed at an accelerating voltage of 3 kV to capture the secondary electron mode images. Semi-quantitative fiber analysis was performed on representative scanning electron microscopy (SEM) images of decellularized uterine tissues acquired at consistent magnification. Fiber diameter was measured using ImageJ (National Institutes of Health, USA) by drawing line measurements perpendicular to clearly identifiable collagen fibers. Fiber orientation was assessed by measuring the angle of individual fibers relative to the horizontal axis. For each analyzed SEM image/group, more than 50 fibers ($n > 50$) were measured. The alignment index was calculated as the axial resultant length of the fiber orientation angles according to the following equation:

$$AI = \left| \frac{1}{n} \sum_{j=1}^{n} e^{i2\theta_j} \right| \tag{2}$$

where $AI$ is the alignment index, $n$ is the number of measured fibers, and $\theta_j$ is the orientation angle of each fiber. The angle was doubled because fiber orientation is axial, meaning that 0° and 180° represent the same direction.

Pore-size analysis of dUECM hydrogels was performed using SEM images acquired at the same magnification. Images were analyzed using ImageJ. Clearly identifiable open pores were selected from representative regions of each image, avoiding image borders, collapsed areas, damaged regions, and preparation-related artifacts. Pores were defined as visible void spaces enclosed or partially enclosed by the ECM network. The apparent pore diameter was measured across the widest internal pore dimension. This analysis was considered semi-quantitative because it was based on two-dimensional SEM images of freeze-dried hydrogels rather than three-dimensional porosity measurements.

#### 2.3.7. Fourier-transform infrared spectroscopy

Fourier-transform infrared (FTIR) spectroscopy was employed to characterize the biochemical composition and relative structural preservation of ECM components in native and decellularized porcine uterine tissues. Semi-quantitative analysis was performed using an internally normalized, ratio-based approach adopted from previously established diamond ATR-FTIR methodologies for collagen structural assessment, in which relative peak areas and intra-band absorbance ratios are used to evaluate molecular organization rather than absolute protein content.[41]

Tissue samples were cryo-ground under liquid nitrogen and sequentially sieved through 600 µm, 300 µm, and 150 µm stainless-steel meshes. The 300 µm fraction was selected for all measurements to ensure uniform particle size, consistent packing density, and reproducible sample presentation across groups. FTIR spectra were acquired using a Spectrum 3 Tri-Range MIR/NIR/FIR FTIR spectrometer (PerkinElmer®, USA) equipped with a diamond attenuated total reflectance (ATR) crystal. The diamond ATR configuration was chosen to minimize sample preparation–induced artifacts, avoid moisture interference associated with KBr pellet methods, and preserve native ECM biochemical signatures. For each measurement, approximately 2 mg of sieved powder was placed onto the ATR crystal and compressed under a constant applied load of 100 N to standardize sample–crystal contact and reduce variability associated with pressure-dependent ATR sampling.

Spectra were collected over the range of 4,000–650 $cm^{-1}$ at a resolution of 4 $cm^{-1}$, using identical acquisition parameters for all samples. All spectra were normalized between 0 and 1 prior to analysis to minimize variability arising from ATR penetration depth, sample thickness, and packing effects. Because the analysis focused on normalized spectra and internal band comparisons, baseline correction was not applied, and FTIR data were interpreted based on relative spectral features rather than absolute absorbance intensities.

Spectral analysis focused on characteristic ECM-associated absorption bands, including the Amide I band (~1,655 $cm^{-1}$), primarily attributed to C=O stretching vibrations of the peptide backbone and sensitive to protein secondary structure; the Amide II band (~1,538 $cm^{-1}$), arising mainly from N–H bending coupled with C–N stretching vibrations; and the Amide III band (~1,234 $cm^{-1}$), involving complex C–N stretching and N–H bending modes associated with collagen structural organization. Additional regions of interest included the hydroxyl and amine stretching region (3,500–3,200 $cm^{-1}$), associated with O–H and N–H vibrations, and aliphatic C–H stretching bands at approximately 2,938 and 2,875 $cm^{-1}$.

Semi-quantitative evaluation of ECM preservation was performed by comparing normalized integrated areas of the Amide I, II, and III bands and by calculating internal absorbance ratios, including Amide III/1,450 $cm^{-1}$ and intraband Amide I ratios (*e.g.*, 1,655/1,690 $cm^{-1}$). Because these ratios compare absorbance features within the same spectrum that are affected equally by ATR sampling conditions, they reduce sensitivity to variations in optical path length and baseline offsets. This internally normalized, ratio-based strategy enables comparative assessment of collagen molecular organization and relative structural preservation across decellularization protocols rather than absolute quantification of collagen or total ECM content.

#### *2.3.8. Raman spectroscopy*

Raman spectroscopy was performed to evaluate the molecular composition of native and decellularized uterine tissues using a Renishaw InVia Reflex Raman microscope (Renishaw Inc., West Dundee, USA). Tissue powders were prepared as described for FTIR analysis, including freeze-drying, cryo-grinding, and sequential sieving, and the 300-µm fraction was used to ensure consistency between the two spectroscopic techniques.

The Raman instrument was equipped with an $Ar^+$ laser (StellarPro-50, Modu-Laser, Centerville) operating at 514.5 nm and an 1,800 lines/mm grating. A Leica 50× NPLAN objective (NA = 0.75) was used to focus the laser onto the sample in backscattering geometry, and scattered photons were collected using a Peltier-cooled CCD detector. Spectra were recorded in extended mode over 100–3,200 $cm^{-1}$ with a detector exposure time of 10 s per scan. Laser power at the sample surface was maintained between 1.8–3.6 mW to avoid thermal damage. The system was calibrated before each measurement session using an internal Si (110) reference peak at 520 $cm^{-1}$.

For each decellularization condition, Raman measurements were acquired from three biological replicates, and three technical spectra were collected from spatially distinct regions of each sample to account for intra-sample heterogeneity and ensure reproducibility. Preprocessing included moving-average smoothing (window size 20), baseline correction, and spectral normalization to enable direct comparison across samples.

Peak fitting and deconvolution were performed using MagicPlot 3.0.1 with standardized parameters applied uniformly across all datasets. Because Raman scattering does not depend on sample contact pressure or optical path length, the technique provided a complementary and methodologically independent confirmation of ECM preservation alongside FTIR analysis.

#### *2.3.9. Thermogravimetric analysis*

To assess the protein content in each decellularization protocol, thermogravimetric analysis (TGA) was performed using a Perkin-Elmer TGA 8,000 analyzer. Approximately 5 mg of the freeze-dried native and decellularized samples underwent heating from ambient temperature to 600 °C at a rate of 10 °C/min under a constant nitrogen gas flow of 30 $cm^3$/min. By differentiating the TGA values, the differential form of TGA (DTA) was derived, facilitating the identification of the maximum disintegration temperature at each stage of thermal degradation.

### 2.4. Hydrogel ink preparation

To prepare the hydrogel ink, dUECM was utilized as the primary component. The decellularization process followed a validated protocol designed to maximize the preservation of essential bioactive molecules and effective removal of cellular material. Once the tissue was decellularized, it was freeze-dried to remove residual moisture, thereby enhancing its stability and shelf-life.

The dried dUECM was subsequently ground into a fine powder to facilitate enzymatic digestion. To achieve this fine particulate form, the freeze-dried tissue was pulverized using dry ice and a grinder to ensure uniformity and minimize aggregation. This step is critical for enhancing the reactivity of the powder in subsequent digestion (Figure 2A).

The required quantity of dUECM powder was weighed and then subjected to enzymatic digestion. To ensure optimal digestion, the Voytik–Harbin method was used.[42] An aliquot of 100 mg of dUECM powder was suspended in a 0.5 M acetic acid solution (Cat.# 351271, Fisherbrand) containing 10 mg pepsin (Cat.# P7125, Sigma-Aldrich) to catalyze the digestion process. The suspension was maintained at a temperature of 4 °C under constant stirring (50 rpm) for 24 h to ensure complete digestion. The cold environment helped preserve the structural integrity of the ECM components during the enzymatic reaction.

Following digestion, the resulting dUECM solution was neutralized with 10 M sodium hydroxide (NaOH, Cat. # BP359, Fisher Scientific).[43] This neutralization step was performed in a refrigerated environment at 4 °C within a walk-in cooling chamber to prevent premature self-crosslinking of collagen molecules, which could compromise the hydrogel's handling and mixing with sodium alginate. The slow and controlled neutralization allowed for the formation of a homogeneous and stable hydrogel suitable for further steps, including cell viability testing and 3D printing.

To ensure the removal of residual components from the neutralization reaction (such as sodium ions and unreacted NaOH) as well as pepsin, the neutralized dUECM solution was subjected to dialysis. A 12–14 kDa molecular weight cut-off (MWCO) membrane was used to retain larger ECM components such as collagen and glycosaminoglycans while allowing smaller molecules like sodium ions, hydroxide ions, and pepsin to diffuse out.

Theoluteion was dialyzed against a neutral PBS buffer. The dialysis process was carried out at 4 °C to prevent degradation of ECM components, with buffer changes performed multiple times over a period of 48 h. This protocol ensured thorough removal of small molecules and enzymes while preserving the bioactivity of the ECM components (Figure 2A). The resulting neutralized and dialyzed dUECM solution was used for rheological characterization and subsequent biological and printing studies.

### 2.5. Rheological characterization and gelation kinetics of dUECM hydrogels

The rheological properties and gelation kinetics of the neutralized dUECM hydrogels were evaluated using a Discovery HR 20 rheometer (TA Instruments®) and a combination of rheometry and turbidimetry tests. For the steady-shear rheology test, viscosity and stress were analyzed as functions of shear rate ($10^{-2}$ to $10^{3}$ $s^{-1}$) at 4 °C. These measurements provided insights into the flow behavior and mechanical stability of the hydrogels under varying flow conditions.

Rheological measurements were conducted using a 60 mm parallel plate geometry. For oscillatory shear test, gelation kinetics were assessed by monitoring the storage modulus (G') and loss modulus (G") across a range of temperatures (10, 15, 20, 25, and 30 °C). The gap size was set to 600 μm with an applied amplitude of 1% strain and a frequency of 1 Hz. This configuration ensured precise thermal control and allowed for accurate simulation of the temperature-dependent gelation behavior of the hydrogels.

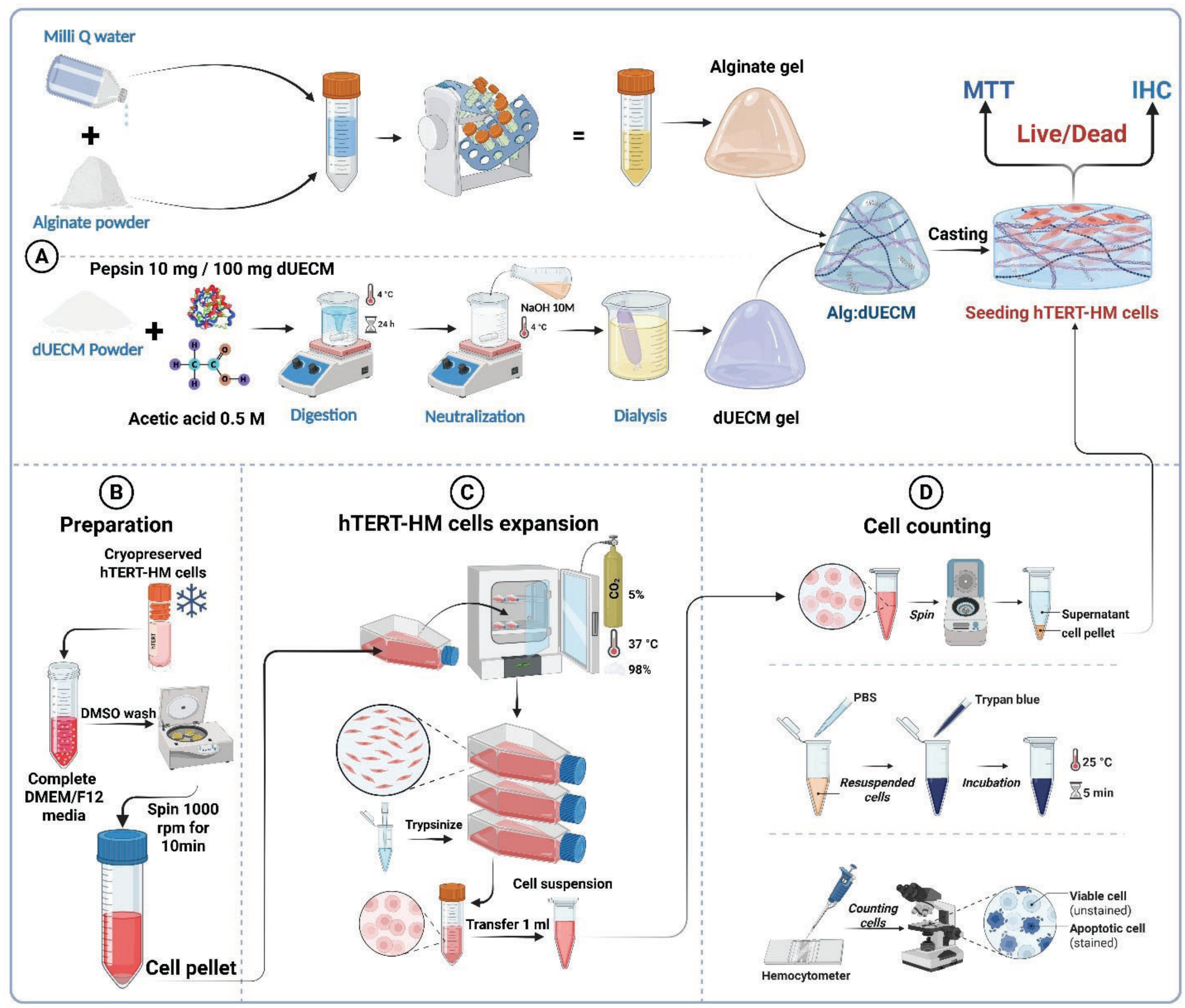

**Figure 2.** Flow diagram of hydrogel ink preparation, hTERT-HM cell expansion, and seeding of the cells for MTT, Live/Dead, and IHC assays
Abbreviations: Alg: Alginate; DMEM: Dulbecco's Modified Eagle Medium; dUECM: Decellularized uterine extracellular matrix; IHC: Immunohistochemistry; PBS: Phosphate-buffered saline.

Gelation kinetics were further investigated using a turbidimetry test conducted at 405 nm. dUECM dispersions at three concentrations (5 mg/mL, 10 mg/mL, and 15 mg/mL) were prepared in triplicate. A BioTek® Synergy HT Multi-Detection Microplate Reader (BioTek® Instruments) was preheated to 37 °C to mimic physiological conditions. For each test, 200 µL of dUECM dispersion was pipetted into a 96-well plate and kept on ice to prevent premature gelation before being transferred to the preheated microplate reader. Absorbance measurements were recorded every 2 min. Changes in absorbance intensity over time were monitored to evaluate gelation progression, including initiation and stabilization times for each concentration. This integrated approach enabled a thorough characterization of the mechanical and kinetic properties of the dUECM hydrogels under physiologically relevant conditions.

### 2.6. Hydrogel ink printability

Hydrogels were prepared using alginic acid sodium salt (Cat.# A2033, Sigma-Aldrich) at two concentrations (2% and 3% w/v) and combined with dUECM at concentrations of 0.5%, 1%, and 1.5% w/v to form the following groups: 2% Alg, 2% Alg + 0.5% dUECM, 2% Alg + 1% dUECM, 2% Alg + 1.5% dUECM, 3% Alg, 3% Alg + 0.5% dUECM, 3% Alg + 1% dUECM, and 3% Alg + 1.5% dUECM. A box-shaped structure (10 mm × 10 mm) was created using 3D builder software in a 3D manufacturing format (3 MF) and sliced into a 25-layer printable structure with a slice thickness defined as 80% of the theoretical strand diameter.

The 3D printing process was carried out using a pneumatic extrusion-based bioprinter (EnvisionTEC® 4th Gen 3D-Bioplotter, Gladbeck) with nitrogen gas at different pressures and speeds depending on printability. The nozzle was pre-warmed for 30 min prior to printing, and the hydrogel was maintained at 25 °C during the printing process. The scaffolds were printed as a 90° lattice structure with a strand infill of 1 mm. After every five layers, a 3-s nozzle purging was performed to prevent clogging. The samples were printed into 12-well plates coated with 0.1% polyethyleneimine (PEI, M.W. 60,000, 50% w/w aq. Thermo Scientific) at 37 °C the day before printing. PEI, a highly positively charged polycation, reacts with the negatively charged carboxyl groups in alginate, forming a polyelectrolyte complex that stabilizes the molecular structure and enhances strand attachment. On the day of printing, the PEI solution was replaced with 50 mM calcium chloride ($CaCl_2$, Cat.# 223506, Sigma-Aldrich) in 0.1% PEI solution, and the scaffolds were printed directly into this solution.

The printed scaffolds were immediately crosslinked in a 100 mM $CaCl_2$ solution for 10 min, washed with Dulbecco's Modified Eagle Medium (DMEM, Cat.# 31600034) cell culture media without fetal bovine serum (FBS), and maintained in complete culture media at 37 °C. To evaluate printability, two-layer constructs were printed, and strand diameters were measured using images captured by an automated fluorescence microscope (BioTek® Lionheart LX) at 4× magnification. ImageJ software (version 1.54 g) was used for the measurements. A total of 50 measurements were performed for each hydrogel group using quadruplicate scaffolds.

Printability was assessed by comparing the measured strand diameters to the theoretical strand diameter (200 µm). Variations in strand diameter were analyzed to determine the impact of hydrogel composition and printing parameters on scaffold fidelity calculated with the following formula:

$$\textbf{\textit{Strand printability}} = 1 - \left( \frac{d_t - d_e}{d_t} \right) \tag{3}$$

where $d_e$ is the diameter of each strand in 3D-printed scaffolds and $d_t$ is the diameter of each strand in the designed one. However, the printability of the hydrogels is usually greater than one due to their high degree of swelling.

### 2.7. Swelling and degradation

To evaluate the swelling and degradation characteristics of 3D-printed alginate-dUECM scaffolds, the samples were incubated in complete Dulbecco's Modified Eagle Medium (DMEM, Cat.# 31600034) culture media containing 1% antibiotic-antimycotic (100×) containing 10,000 units/mL penicillin, 10,000 µg/mL of streptomycin at 37 °C under 5% $CO_2$. The scaffolds were fabricated under sterile conditions to prevent contamination that might influence swelling and mass loss measurements.

Immediately after printing, the scaffolds were gently blotted with sterile Kimwipes™ to remove excess moisture, and their initial wet weight ($W_0$) was recorded. The scaffolds were then immersed in complete cell culture media and incubated at 37 °C with 5% $CO_2$. At predetermined intervals (0, 1, 3, 7, and 14 days), the scaffolds were removed from the media, blotted, and their wet weight ($W_t$) was measured. The swelling percentage was calculated using the formula:[44]

$$\textbf{\textit{Swelling}}\,(\%) = \frac{W_t - W_0}{W_0} \times 100 \tag{4}$$

where $W_t$ represents the wet weight of the scaffold at each time point, and $W_0$ is the initial wet weight. Measurements were conducted in quintuplicate for each time point to ensure statistical accuracy.

Following the swelling assessments, the scaffolds were freeze-dried to determine their dry weight at specific intervals ($W_t$). Separate scaffolds were lyophilized immediately after printing to establish the initial dry weight ($W_0$). The remaining mass percentage was calculated using the formula:[44]

$$\textit{Remaining Mass}(\%) = \frac{W_0 - W_l}{W_0} \times 100 \quad (5)$$

where $W_0$ is the initial dry weight of the scaffold, and $W_l$ is the lyophilized weight at each time point. This process was performed for each interval (0, 1, 3, 7, and 14 days) using separate sets of scaffolds to maintain accuracy. All experiments were conducted in sterile conditions, and results were analyzed to determine the swelling and degradation behaviors of the scaffolds over time.

### 2.8. Cell viability, metabolic activity, and immunocytochemical evaluation of developed hydrogel inks

Cell viability of the optimal hydrogel ink was assessed using MTT (3-[4,5-dimethylthiazol-2-yl]-2,5-diphenyltetrazolium bromide, Cat. # M6494, Invitrogen) and Live/Dead, and immunocytochemical (IHC) assays. Prepared Alg:dUECM hydrogels were cast into 24-well plates, and $5 \times 10^4$ hTERT-HM cells[45] were seeded on the selected hydrogel surfaces ($n$ = 6). Cell-seeded constructs were incubated under standard culture conditions at 37 °C with 5% $CO_2$ and were maintained in complete culture media including, DMEM/F12 (Dulbecco's Modified Eagle Medium/Nutrient Mixture F-12, HEPES, Cat. # 11330-032, Gibco) supplemented with 10% FBS (Gibco®, Cat. # A5256701) and 1% antibiotic-antimycotic (100×) containing 10,000 units/mL penicillin, 10,000 µg/mL of streptomycin analyzed at predetermined time points (days 1, 3, 5, and 7) (Figure 2B–D).

**MTT assay:** To evaluate the metabolic activity of hTERT-HM cells seeded on the hydrogels, MTT assays were performed on days 1, 3, 5, and 7. MTT solution (0.5 mg/mL) was added to each well and incubated for 4 h at 37 °C, allowing metabolically active cells to form formazan crystals. These crystals were dissolved in dimethyl sulfoxide (DMSO, Cat. # BP231-1, Fisher bioreagents), and absorbance was measured at 550 nm using a microplate reader (BioTek Synergy HT Multi-Detection Microplate Reader) to assess cell proliferation and viability over time. Since most formazan crystals formed within the hydrogel matrix and DMSO alone made the casted gel rigid, hindering efficient dye elution, the hydrogels were crushed using a tissue grinder (Bio-Gen PRO200 Homogenizer) to facilitate dye extraction, which was then isolated via centrifugation. This step facilitated effective dye elution, enabling accurate optical density (OD) measurement.

**Live/Dead cell viability assay:** The live/dead assay was performed to evaluate the viability, attachment, and distribution of hTERT-HM cells on the optimal Alg:dUECM hydrogel composition. Calcein AM (Cat. # AS-89201, AnaSpec, Fremont) and propidium iodide (PI, Cat. # AS-83215, AnaSpec, Fremont) were used to stain live and dead cells, respectively. Imaging was performed using a Lionheart Biotek fluorescence microscope in stacked imaging mode to minimize blurring caused by surface irregularities. The images were deconvoluted using the microscope's software to enhance clarity and analyze cell viability and morphology.

**Immunohistochemical (IHC) staining for α-smooth muscle actin (α-SMA):** To evaluate cytoskeletal organization and uterine-relevant cellular phenotype, immunocytochemical staining for α-smooth muscle actin (α-SMA; ACTA2) was performed on cell-seeded scaffolds at days 1, 3, 5, and 7. Constructs were fixed in 4% paraformaldehyde (PFA) in PBS, washed thoroughly with PBS, and stored at 4 °C until staining. Prior to antibody incubation, scaffolds were washed in PBS for 5 min and permeabilized with PBS containing 0.1% Triton X-100 (PBT) for 15 min at room temperature on a shaker. After an additional PBS wash, nonspecific binding was blocked for 30 min at room temperature using a blocking solution composed of 5% goat serum (Vector Laboratories, Cat. # VECTS1000), 1% horse serum, and 1% FBS in PBS. Scaffolds were then incubated overnight at 4 °C with Anti-α-Smooth Muscle Actin (ACTA2) antibody (Sigma-Aldrich, Cat. #A2547) diluted in blocking buffer. Samples were placed face-down on 200 µL droplets of antibody solution on Parafilm within a humidified chamber to ensure uniform antibody contact with the scaffold surface.

Following primary antibody incubation, scaffolds were washed three times with PBT and incubated for 1 h at room temperature in the dark with a rhodamine-conjugated secondary antibody (Sigma-Aldrich, Cat. #AP124R) diluted 1:150 in blocking buffer. After secondary antibody incubation, samples were washed three times with PBT and twice with PBS (5 min each). Cell nuclei were counterstained with 4',6-diamidino-2-phenylindole (DAPI, Sigma-Aldrich, Cat. #D9542) at a dilution of 1:1,000 in PBS for 5 min at room temperature, followed by a final PBS wash. Scaffolds were stored in PBS at 4 °C, protected from light, until imaging. Fluorescence images were acquired using a Nikon Eclipse Ti2-E fluorescence microscope under identical acquisition settings for all experimental groups.

### 2.9. Mechanical properties

#### *2.9.1.Tensile testing of decellularized tissues*

Tensile testing was performed to evaluate the macroscopic mechanical behavior of native and decellularized uterine tissues. The myometrial layer was isolated by carefully removing the endometrium, and tissues were processed using four decellularization protocols with treatment durations of 48 h and 72 h. Native myometrium served as the control group.

Following decellularization and the final washing step, tissues were used directly for mechanical testing. Tissue sheets were cut transversely to obtain membrane-shaped samples. A standardized cutting template based on ISO 527-2:2012 was used to produce uniform "dog-bone" specimens (Figure S1), minimizing stress concentrations and ensuring reproducible dimensions. A 3D-printed support system adapted from Scholze *et al.*[46] was used to stabilize the tissue during cutting and maintain consistent sample thickness.

All samples were visually inspected, and any irregularities were trimmed to comply with ISO dimensional guidelines. Samples were kept hydrated in PBS during handling and immediately mounted in the testing apparatus following the final wash.

Mechanical characterization was performed using a Mark-10 ESM303 uniaxial tensile testing system (Mark-10 Corporation, USA) equipped with a 50 N load cell. Care was taken to align specimens axially and prevent gripping-induced damage. Tests were conducted at room temperature with a constant crosshead speed of 0.01 mm/s, appropriate for soft biological tissues.

For each specimen, a stress–strain curve was generated, and the following mechanical parameters were calculated: ultimate tensile strength (UTS), Young's modulus (initial linear region), and elongation at break (%). Each decellularization protocol and treatment duration was tested using three biological replicates ($n$ = 3 per group). These measurements were used to compare differences in tissue-level mechanical behavior following various decellularization conditions.

#### *2.9.2. Compressive test on scaffolds*

To determine the elastic modulus of the 3D-printed scaffolds, a uniaxial compression test was conducted using the ElectroForce® BioDynamic® 5100 Series device equipped with a 20 N load cell. Prior to testing, each scaffold dimension (x, y, and z) was carefully measured using a digital calliper.

The scaffolds were positioned on the test platform to ensure proper alignment and avoid uneven loading. Compression was applied at a constant speed of 0.01 mm/s, and the scaffolds were compressed to 50% of their original height. Throughout the compression process, the device recorded stress-strain data in real-time, generating stress-strain curves for each sample.

The elastic modulus was calculated from the slope of the linear region of the stress-strain curve, corresponding to the material's elastic deformation phase. The experiment was repeated for at least three samples from each scaffold group to ensure reproducibility and statistical significance of the results.

### 2.10. Statistical analysis

Data were analyzed using IBM SPSS Statistics software version 27.0 (IBM Corp., Armonk, NY, USA). Quantitative variables are reported as mean ± standard deviation (SD). The normality of data distribution was assessed using the Shapiro–Wilk test, and appropriate statistical tests were selected accordingly. One-way ANOVA was conducted to compare group means. For post-hoc pairwise comparisons, Dunnett's test was employed to compare each experimental group with the designated control group. Statistical significance was defined as follows: **** for $p < 0.0001$, *** for $p < 0.001$, ** for $p < 0.01$, * for $p < 0.05$, and NS for non-significant differences.

## 3. Results

### 3.1. Preparation and characterization of uterine-derived dECM

The first objective of this study was to establish an efficient decellularization strategy for porcine uterine tissue that effectively removes cellular components while preserving ECM. Decellularization efficacy and ECM preservation were comprehensively evaluated using biochemical, histological, spectroscopic, and mechanical analyses.

#### *3.1.1. Quantitative assessment of residual DNA content and fragment size*

The native tissue's DNA content was 9,142.67 ± 26.63 ng/mg dry weight, thus serving as the control. Figure 3A shows that treatments combining 1% Triton™ X-100 and SDS at various concentrations and durations significantly decreased DNA content; each treatment's percentage reduction was calculated.

DNA content decreased progressively with increasing SDS concentration (Figure 3A). At 0.1% SDS, residual DNA remained high (419.33 ± 9.45 and 408.67 ± 2.31 ng/mg at 48 and 72 h, respectively; ~95% reduction), indicating incomplete decellularization. Treatment with 0.5% SDS achieved greater removal (82.67 ± 5.03 ng/mg

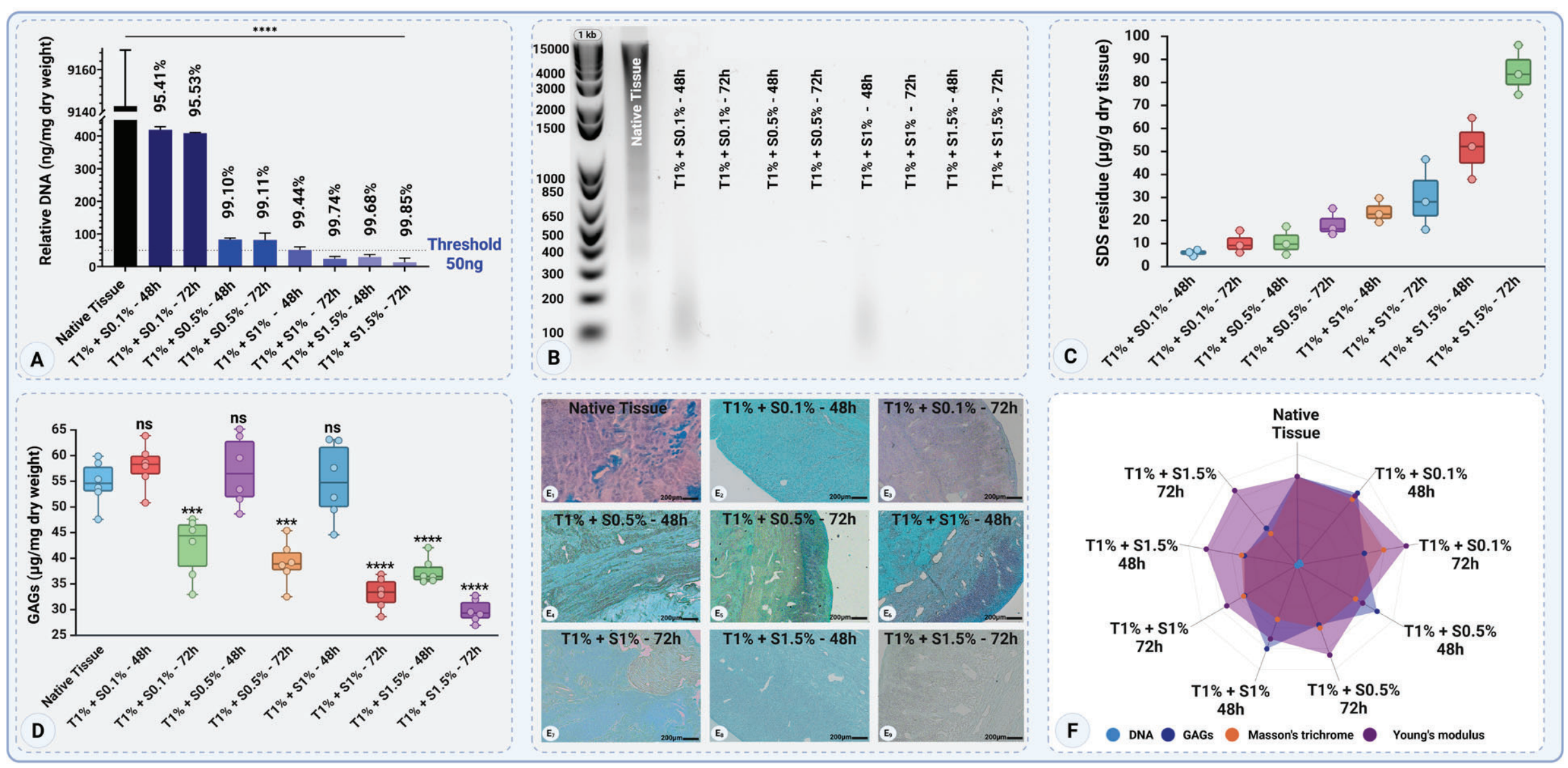

**Figure 3.** Quantitative assessment of decellularization efficiency. (A) DNA quantification shows a substantial reduction of DNA across all protocols relative to native tissue. (B) Gel electrophoresis confirms progressive DNA fragmentation and complete loss of detectable bands in SDS ≥0.5% groups. (C) SDS residue measured by MBAS assay, demonstrating low detergent retention in mild protocols and increasing residue at higher SDS concentrations. (D) Biochemical GAG quantification (mean ± SD) aligns with histology, showing higher retention in low-SDS groups and depletion at higher SDS concentrations or longer exposure. ($E_1$–$E_9$) Alcian blue staining illustrates GAG distribution and preservation patterns across treatment conditions. Scale bar: 200 µm. (F) Radar plot integrating DNA, GAGs, collagen staining score, and mechanical properties to compare overall protocol performance. Statistical significance: ns = not significant; $p < 0.05$ (*), $p < 0.01$ (**), $p < 0.001$ (***), $p < 0.0001$ (****). Data are expressed as mean ± SD.

at 48 h; ~99% reduction), though still above the 50 ng/mg threshold.

At 1% SDS, residual DNA was 51.33 ± 9.02 ng/mg at 48 h (99.44% reduction) and 24 ± 7.21 ng/mg at 72 h (99.74% reduction). The 1.5% SDS condition achieved the greatest reduction, yielding 29.33 ± 7.57 ng/mg at 48 h and 13.33 ± 12.86 ng/mg at 72 h (99.85% reduction). Full numerical data for all groups are presented in Figure 3A.

All treatments with SDS concentrations of 1% or higher achieved substantial DNA removal, with Triton™ 1% + SDS 1.5% at 72 h demonstrating the most effective DNA elimination. These findings highlight the efficiency of combining Triton™ 1% with SDS for decellularization, where higher SDS concentrations and extended treatment durations result in superior DNA elimination (Figure 3A). It is noted that only the T1% + SDS 1% at 72 h, and T1% + SDS 1.5% at both 48-h and 72-h groups achieved DNA values strictly below the commonly accepted threshold of 50 ng/mg dry weight. The T1% + SDS 1% at 48-h group yielded 51.33 ± 9.02 ng/mg, which marginally exceeds this threshold. However, gel electrophoresis (Figure 3B) confirmed the complete absence of detectable high-molecular-weight DNA fragments in this group, indicating that residual DNA is highly fragmented and unlikely to elicit a significant immunogenic response. Protocol selection was therefore based on a multi-parameter assessment encompassing DNA content, GAG retention, collagen structural integrity, and mechanical properties, the rationale for which is detailed in the protocol selection section below (Figure 3A).

All treatments with SDS concentrations of 1% or higher achieved substantial DNA removal, with Triton™ 1% + SDS 1.5% at 72 h demonstrating the most effective DNA removal. Gel electrophoresis (Figure 3B) showed intact high-molecular-weight DNA in native tissue, partial fragmentation at 100–400 bp with 0.1% SDS, and complete absence of detectable bands at ≥0.5% SDS. The T1% + SDS 1% at 48-h group yielded 51.33 ± 9.02 ng/mg dry weight—marginally above the commonly accepted 50 ng/mg threshold—yet showed no detectable high-molecular-weight fragments, indicating highly fragmented, immunologically low-risk residual DNA. Protocol selection was therefore based on a multi-parameter assessment of DNA content, GAG retention, collagen integrity, and mechanical properties (Figure 3A), detailed in the protocol selection section below.

#### 3.1.2. SDS residue quantification (MBAS assay)

Residual SDS content increased in a concentration-dependent manner across protocols (Figure 3C). The lowest detergent levels were observed in tissues treated with 0.1 and 0.5% SDS, with residues of 6.02 ± 1.37 μg/g for T1% + S0.1% (48 h), 10.31 ± 4.90 μg/g for T1% + S0.1% (72 h), 10.80 ± 6.16 μg/g for T1% + S0.5% (48 h), and 18.60 ± 5.87 μg/g for T1% + S0.5% (72 h). These values reflect efficient detergent removal at low SDS concentrations. Increasing SDS to 1% resulted in higher residues of 23.92 ± 5.33 μg/g at 48 h and 30.26 ± 15.35 μg/g at 72 h, indicating greater detergent interaction with the ECM. The highest SDS retention occurred in the 1.5% SDS groups, with 51.54 ± 13.34 μg/g at 48 h and 84.80 ± 10.82 μg/g at 72 h, representing a substantial increase in residual detergent. Overall, SDS retention followed a concentration-dependent pattern, with extended incubation enhancing detergent retention predominantly at 1.5% SDS.

#### 3.1.3. GAG quantification and alcian blue staining

Native porcine uterine tissue contained 54.68 ± 4.38 μg/mg dry weight of GAGs (Figure 3D). GAG retention was inversely related to SDS concentration and exposure duration. Mild conditions (≤0.5% SDS, 48 h) preserved or slightly exceeded native GAG levels (57.92 and 57.02 μg/mg, respectively), while treatment for 72 h at the same concentrations resulted in notable depletion (42.17 and 39.16 μg/mg).

At 1% SDS for 48 h, GAG content was preserved at near-native levels (54.94 ± 7.55 μg/mg), but extended to 72 h resulted in marked depletion (33.19 ± 3.09 μg/mg). The 1.5% SDS groups exhibited the greatest GAG loss across all conditions (37.51 and 29.75 μg/mg at 48 and 72 h, respectively). Collectively, SDS concentration and exposure duration were the primary determinants of GAG retention, with shorter, milder treatments yielding superior ECM preservation (Figure 3D). Alcian Blue staining supported these quantitative findings (Figure 3$E_1$–$E_9$). Native tissue exhibited intense, uniform blue coloration, reflecting abundant sulfated GAGs. Mild decellularization protocols (0.1% and 0.5% SDS for 48 h) retained strong staining comparable to native tissue, whereas higher SDS concentrations and longer exposure times produced visibly diminished staining intensity. The most aggressive treatment (1.5% SDS for 72 h) showed markedly weak staining, correlating with the lowest GAG levels measured biochemically.

Together, these results demonstrate that SDS concentration and exposure duration have significant effects on GAG retention, with milder conditions yielding superior preservation of ECM polysaccharides.

To visualize the combined effects of decellularization conditions on biochemical and mechanical outcomes, a multivariate correlation plot was generated (Figure 3F). This analysis highlights the relationship between DNA removal efficiency, GAG retention, and Young's modulus across different detergent concentrations and exposure times. Conditions associated with higher SDS concentrations and longer treatment durations demonstrated more effective DNA removal but were accompanied by greater reductions in GAG content and altered mechanical properties. In contrast, intermediate conditions showed a more balanced profile, maintaining ECM components while achieving sufficient decellularization. This integrative comparison supports the selection of optimized decellularization parameters that balance biochemical preservation with mechanical integrity.

#### 3.1.4. Histological assessment

Figure 4 illustrates the native porcine uterine tissue, showcasing the three distinct layers: the endometrial layer, the inner circular smooth muscle layer of the myometrium, and the outer longitudinal smooth muscle layer of the myometrium.[47] The structural organization of each layer is shown using DAPI staining (Figure 4$A_1$–$A_3$), schematic representations (Figure 4$B_1$–$B_3$), and H&E staining (Figure 4$C_1$–$C_3$).

The endometrium (Figure 4$A_1$) shows dense nuclear localization under DAPI staining, supported by the schematic (Figure 4$B_1$) illustrating glandular and vascular structures, and H&E staining (Figure 4$C_1$) confirming the presence of blood vessels and glands. The inner circular smooth muscle layer (Figure 4$A_2$) is marked by evenly distributed nuclei in DAPI staining, with the schematic (Figure 4$B_2$) and H&E staining (Figure 4$C_2$) highlighting its dense, circularly arranged smooth muscle fibers. The outer longitudinal smooth muscle layer (Figure 4$A_3$) displays sparsely distributed nuclei in DAPI staining, with schematic (Figure 4$B_3$) and H&E staining (Figure 4$C_3$) emphasizing its longitudinal alignment of muscle fibers.

Additionally, SEM images provide detailed views of the tissue's surface and cross-sectional features. The surface of the endometrium (Figure 4$D_1$) reveals microvilli structure, indicative of its functional role in nutrient exchange and secretion, while the cross-sectional view (Figure 4$D_2$) highlights its vascularity with visible red blood cells (RBCs). The serosa, or peritoneal surface, is shown in Figure 4$E_1$ with its smooth texture and folds, while Figure 4$E_2$ provides a cross-sectional view of the myometrium, displaying fibrous structures with visible RBCs embedded. These findings provide a comprehensive overview of the native uterine tissue's organization and functional features.

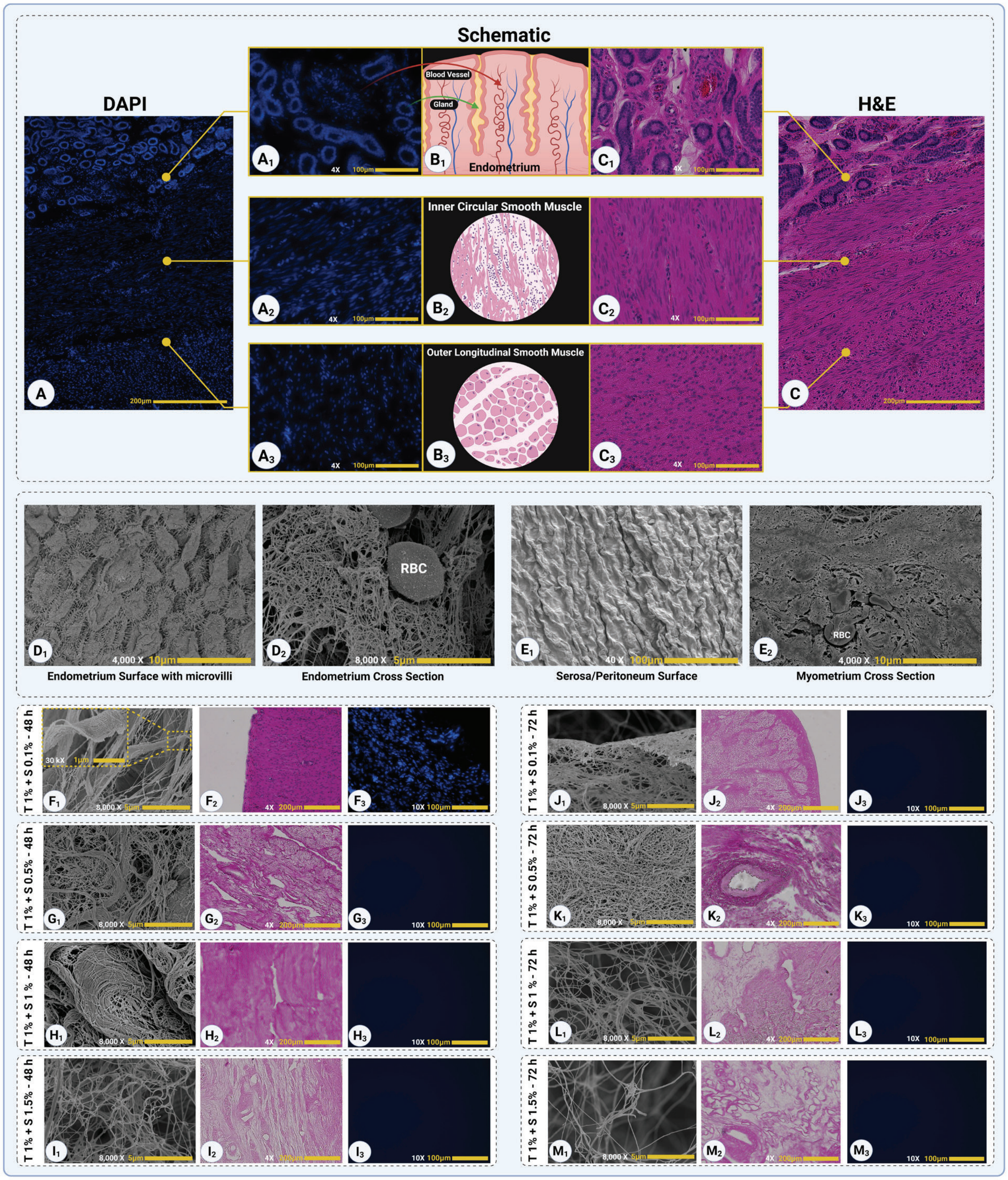

**Figure 4.** Structural characterization of native and decellularized porcine uterine tissue. Panels A–E depict the multiscale architecture of native porcine uterus, including histological organization and ultrastructural features assessed by H&E staining, DAPI nuclear staining, and scanning electron microscopy (SEM). Panels F–M show SEM (F1–M1), H&E (F2–M2), and DAPI (F3–M3) images of uterine tissue following decellularization using Triton™ X-100 (1%) combined with SDS at concentrations of 0.1%, 0.5%, 1%, and 1.5% for 48 h and 72 h. All H&E images were acquired at 4× magnification and DAPI images at 10× magnification using identical exposure settings across all groups. Scale bars are indicated on individual panels. All SEM images of decellularized tissues (F1–M1) are presented at 8,000× magnification to facilitate direct comparison of ECM ultrastructure across treatment groups; scale bars are indicated in each panel. Native tissue SEM panels are shown at feature-specific magnifications to illustrate anatomical surface and cross-sectional characteristics of untreated uterine tissue.

The decellularized samples were analyzed to assess cellular removal and preservation of ECM architecture. In the Triton™ X-100 1% + SDS 0.1% 48 h group, residual nuclei were still visible in dense tissue regions, predominantly within the myometrium, as observed by both H&E (Figure 4F$_2$) and DAPI staining (Figure 4F$_3$). SEM analysis further showed that the collagen fibrillar network was less distinctly exposed compared with groups treated with higher SDS concentrations (Figure 4F$_1$ and Table S1). The ECM surface appeared partially covered by residual cellular or matrix-associated material, with collagen fibrils embedded within or masked by sheet-like residual matrix components. Therefore, an additional higher-magnification SEM image was included for this group to better demonstrate the underlying fibrillar architecture. Together, these findings indicate that the mildest detergent condition was insufficient for complete decellularization, consistent with the biochemical DNA results. For the remaining 48-h treatment groups (Triton™ 1% + SDS 0.5%, 1%, and 1.5%), no visible nuclei were detected in either H&E (Figure 4G$_2$, H$_2$, and I$_2$, respectively) or DAPI (Figure 4G$_3$, CH$_3$, and I$_3$) staining, indicating effective cellular removal. The SEM images (Figure 4F$_1$–I$_1$) of these groups revealed clear collagen fibers, demonstrating the preservation of the ECM's fibrous structure.

In Triton™ 1% + SDS 0.1% (72 h), collagen fibrils became visible in SEM (Figure 4J$_1$), although residual non-collagenous matrix material was still present on the fibrils. Despite these residual materials, no visible nuclei were observed in H&E (Figure 4J$_2$) or DAPI (Figure 4J$_3$) staining, indicating that extended treatment duration effectively removed cellular components. For the other 72-h treatments (Triton™ 1% + SDS 0.5%, 1%, and 1.5%), no nuclei were detected in H&E or DAPI staining, and SEM revealed well-defined collagen fibrils (Figure 4K$_{1-3}$, L$_{1-3}$, M$_{1-3}$), confirming complete decellularization and retention of the ECM. SEM analysis revealed distinct differences in collagen fiber organization among the treatment groups. In samples treated with Triton™ 1% + SDS 0.5% and 1% for 48 h, collagen fibrils remained relatively well-organized and preserved (Figure 4G$_1$ and H$_1$), indicating limited structural disruption of the ECM. In contrast, the Triton™ 1% + SDS 1.5% for the 48-h treatment (Figure 4I$_1$) exhibited more disordered and fragmented collagen arrangements, reflecting greater alterations to the ECM microarchitecture.

In the 72-h treatment groups, differences in collagen microarchitecture became more pronounced as a function of SDS concentration. The Triton™ 1% + SDS 0.1% (72 h) condition preserved a dense and highly interconnected collagen fibrillar network, with fine, continuous fibers and minimal evidence of structural disruption (Figure 4J$_1$). However, despite this favorable ultrastructural preservation, residual cellular material is visible.

The Triton™ 1% + SDS 0.5% (72 h) protocol exhibited a well-organized and uniformly aligned collagen fibrillar structure (Figure 4K$_1$), reflecting an optimal balance between effective cellular removal and extracellular matrix (ECM) preservation. Collagen fibers remained continuous and anisotropically arranged, with reduced evidence of fiber fragmentation or collapse.

In contrast, increasing SDS concentration to 1% and 1.5% resulted in progressive disruption of ECM architecture. Samples treated with Triton™ 1% + SDS 1% (72 h) showed less organized and more heterogeneous collagen networks, characterized by irregular fiber orientation and localized thinning (Figure 4L$_1$). The Triton™ 1% + SDS 1.5% (72 h) condition further exacerbated these effects, producing fragmented, loosely packed collagen fibers and regions of apparent matrix degradation (Figure 4M$_1$), consistent with excessive detergent-induced ECM damage.

#### *3.1.5. Collagen preservation and microstructural organization assessed by Masson's trichrome and Picrosirius Red staining*

Masson's trichrome staining was employed to qualitatively assess collagen preservation and overall tissue composition following decellularization, while Picrosirius Red staining under polarized light was used as a complementary technique to evaluate collagen fiber organization and the presence of non-collagenous tissue components (Figure 5).

Native uterine tissue exhibited intense and continuous blue collagen staining with Masson's trichrome, reflecting the dense extracellular matrix characteristic of intact uterine architecture. Prominent, red-stained regions corresponding to muscle and cellular components were also observed, consistent with the high cellularity of native tissue (Figure 5A$_1$; Figures S2 and S3). Under Picrosirius Red staining, native tissue displayed collagen fibers with strong red birefringence, indicating well-organized collagen bundles. In addition, distinct grey regions were evident, corresponding to smooth muscle cells, which are known to exhibit weak or non-birefringent signals under polarized light (Figure 5B$_1$)

After 48 h of decellularization, collagen preservation was strongly dependent on SDS concentration. Tissue treated with Triton™ 1% + SDS 0.1% (48 h) retained substantial blue collagen staining, indicating minimal collagen loss; however, residual red-stained non-collagenous regions remained visible, suggesting incomplete removal of cellular components under these mild conditions (Figure 5A$_2$). Consistently, Picrosirius Red staining of this group showed red birefringent collagen fibers alongside grey regions attributable to residual smooth muscle cells, indicating partial preservation of non-collagenous tissue elements (Figure 5B$_2$).

The Triton™ 1% + SDS 0.5% (48 h) condition demonstrated preserved collagen architecture with blue-stained regions (Figure 5A$_3$). Picrosirius Red staining of this condition was characterized predominantly by red birefringence, with black backgrounds, indicating effective removal of smooth muscle cells while maintaining collagen fiber structure (Figure 5B$_3$).

In contrast, tissues treated with higher SDS concentrations (1% and 1.5%) for 48 h exhibited diminished blue collagen staining and disrupted matrix continuity, consistent with detergent-induced collagen damage (Figure 5A$_4$ and 5A$_5$). Picrosirius Red staining of these groups showed red birefringent collagen fibers with fragmented and heterogeneous patterns, reflecting extensive removal of cellular components and compromised collagen organization (Figure 5B$_4$ and 5B$_5$).

Extending the decellularization duration to 72 h further accentuated these effects across all SDS concentrations. Samples treated with lower SDS concentrations showed reduced collagen density compared to their 48-h counterparts but largely maintained overall matrix continuity. However, tissues exposed to higher SDS concentrations (1% and 1.5%) for 72 hours exhibited pronounced collagen depletion and loss of native-like tissue organization (Figure 5A$_6$ and 5A$_9$). Picrosirius Red staining of these groups continued to demonstrate red birefringence without detectable gray regions, consistent with extensive collagen remodeling and complete removal of smooth muscle cells (Figure 5B$_6$ and 5B$_9$).

#### 3.1.6. FTIR spectral analysis

ATR-FTIR spectroscopy was employed to evaluate relative changes in ECM-associated chemical functionalities following decellularization using Triton™ X-100 and SDS (Figure 6). All spectra were normalized to enable qualitative and semi-quantitative comparison of band shape and relative band ratios across treatment conditions.

The Amide II band (~1,538 cm$^{-1}$) arises predominantly from in-plane N–H bending coupled with C–N stretching vibrations, providing insight into the hydrogen-bonding environment of ECM proteins. While present in all samples, this band exhibited gradual attenuation in protocols employing higher SDS concentrations (1–1.5%) and extended treatment durations (72 h), suggesting progressive disruption of intermolecular hydrogen bonding within the collagen network (Figure 6A).

The Amide III band (~1,234 cm$^{-1}$), which involves complex N–H bending and C–N stretching vibrations, is closely associated with collagen secondary structure and triple-helix integrity. Compared with native tissue,

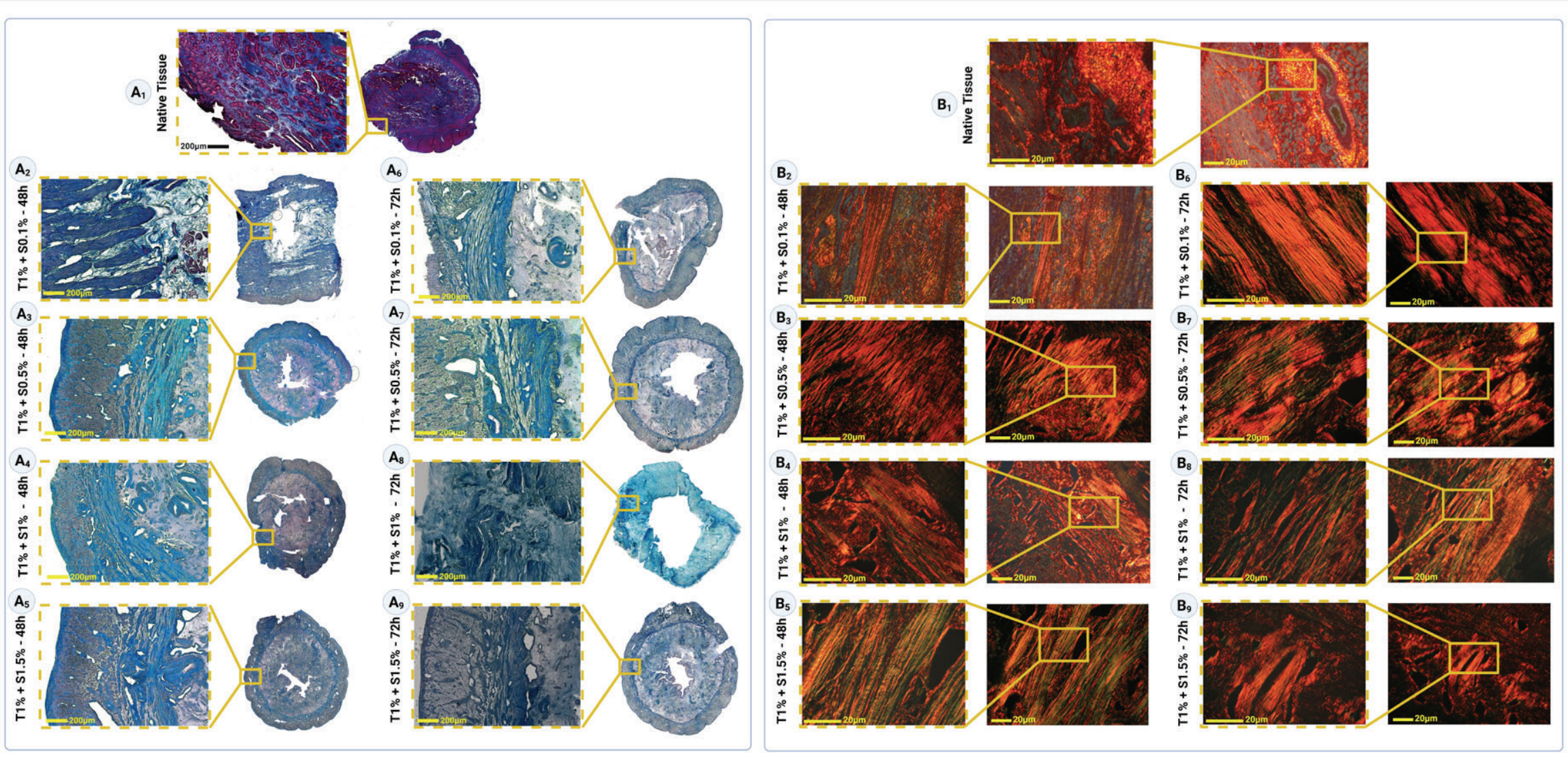

**Figure 5.** Masson's trichrome and Picrosirius Red staining of native and decellularized porcine uterine tissue. (A$_1$–A$_9$) Masson's trichrome–stained sections showing collagen distribution and overall tissue composition in native tissue (A$_1$) and tissues decellularized using Triton™ X-100 (1%) combined with SDS at concentrations of 0.1%, 0.5%, 1%, and 1.5% for 48 h (A$_2$–A$_5$) and 72 h (A$_6$–A$_9$). (B$_1$–B$_9$) Corresponding Picrosirius Red–stained sections imaged under polarized light, illustrating collagen fiber organization and the presence of smooth muscle regions across the same native and decellularized conditions.

Amide III band definition was better preserved in milder decellularization conditions, particularly T1% + S0.1% and T1% + S0.5% at 48 h, whereas higher SDS concentrations and longer exposure times resulted in visibly reduced band intensity and increased spectral smoothing (Figure 6A).

In addition to amide-associated regions, the aliphatic C–H stretching bands (~2,938 and 2,875 $cm^{-1}$) remained relatively consistent across all protocols, indicating minimal disruption to lipid-associated chemical groups despite detergent treatment. In contrast, the broad hydroxyl (–OH) and amine (–NH) stretching region (3,500–3,200 $cm^{-1}$) showed progressive attenuation with increasing SDS concentration and treatment duration. This reduction was most pronounced in protocols employing 1–1.5% SDS for 72 h, consistent with the loss of hydroxyl-containing ECM components and disruption of hydrogen-bond–rich molecular environments (Figure 6A).

The Amide III/1,450 ratio, a critical marker of collagen triple helix integrity, provides valuable insights into the impact of decellularization protocols (Figure 6B). With a threshold of 1 representing preserved collagen structure, the analysis revealed notable variations across different treatment conditions. More corresponding peak wavenumbers, chemical assignments, vibration types, and structural significance are given in Table S2.

The native tissue exhibited a mean ratio of 0.9915 ± 0.02545, which is close to the threshold, indicating a well-preserved collagen structure in its natural state. Among the decellularized groups, the T1% + S0.1% - 48h protocol showed a slight increase in the ratio to 1.0390 ± 0.05766, surpassing the threshold and suggesting minimal structural disruption. In contrast, the T1% + S0.1% - 72h group decreased to 0.9729 ± 0.04485, reflecting a mild collagen degradation with extended treatment duration.

The T1% + S0.5% - 48h protocol yielded a ratio of 1.0136 ± 0.02026, indicating preserved collagen structure. However, the T1% + S0.5%-72h group showed a substantially reduced ratio to 0.9263 ± 0.06783, highlighting the detrimental effect of prolonged exposure at this detergent concentration. Similarly, the T1% + S1% - 48h protocol exhibited the highest ratio of 1.0452 ± 0.03107 among all groups, reflecting excellent collagen preservation despite the increased detergent concentration. Extending the treatment to T1% + S1% - 72h led to a slight decrease in the ratio to 0.9797 ± 0.01149, though it remained near the threshold, indicating moderate preservation.

The T1% + S1.5% - 48h protocol maintained a ratio of 1.0015 ± 0.03287, effectively preserving collagen structure. However, the T1% + S1.5% - 72h protocol showed a slight decline to 0.9992 ± 0.033610, demonstrating near-threshold preservation with evidence of mild degradation.

The 1655/1690 ratio indicates collagen crosslinking and secondary structure integrity (Figure 6C). A higher ratio suggests better preservation of collagen's native triple-helical structure, while lower values indicate structural alterations. The provided results reveal significant variations across different treatment protocols, reflecting the effects of decellularization conditions on collagen integrity.

The native tissue exhibited a mean ratio of 2.7631 ± 0.3194, indicating well-preserved collagen crosslinking and secondary structure in its unaltered state. Among the decellularized groups, the T1% + S0.1% - 48h protocol reduced to 2.3739 ± 0.2525, reflecting moderate structural disruption. Interestingly, the T1% + S0.1% - 72h protocol demonstrated a ratio recovery of 2.6360 ± 0.01975, suggesting that extended exposure under low detergent concentration might stabilize collagen crosslinking in certain conditions.

For the T1% + S0.5% - 48h protocol, the ratio was relatively high at 2.6973 ± 0.1748, indicating strong preservation of collagen integrity during shorter treatments with moderate detergent concentration. However, the T1% + S0.5% - 72h group experienced a notable decrease in the ratio to 2.3661 ± 0.09048, highlighting the adverse impact of prolonged exposure.

The T1% + S1% - 48h protocol maintained a ratio of 2.7164 ± 0.2228, demonstrating effective preservation of collagen structure under higher detergent concentrations over shorter durations. In contrast, the T1% + S1% -72h group exhibited a significant decline in the ratio to 2.3319 ± 0.05083, indicating pronounced structural disruption with extended treatment. Similarly, the T1% + S1.5% - 48h protocol resulted in a ratio of 2.3490 ± 0.11610, while the T1% + S1.5% - 72h group showed a partial recovery to 2.5372 ± 0.02916, suggesting some stabilization in collagen crosslinking under these conditions.

Studies indicate that the 1,655/1,690 $cm^{-1}$ ratio is sensitive to collagen secondary structure and crosslinking organization, and may decrease with increasing detergent concentration or prolonged chemical exposure, consistent with detergent-induced alteration of collagen molecular structure.[41,48] In the present study, statistical analysis using one-way ANOVA followed by Dunnett's multiple comparisons test was performed against native tissue. Groups that did not show a significant difference from native tissue were interpreted as retaining collagen-related molecular features comparable to the native ECM, whereas groups showing significant reductions reflected greater detergent-associated structural alteration. The partial recovery observed in some groups may reflect retention or reorganization of collagen-associated molecular interactions under specific detergent conditions, as previously reported for lower detergent concentrations or less disruptive processing conditions.[49]

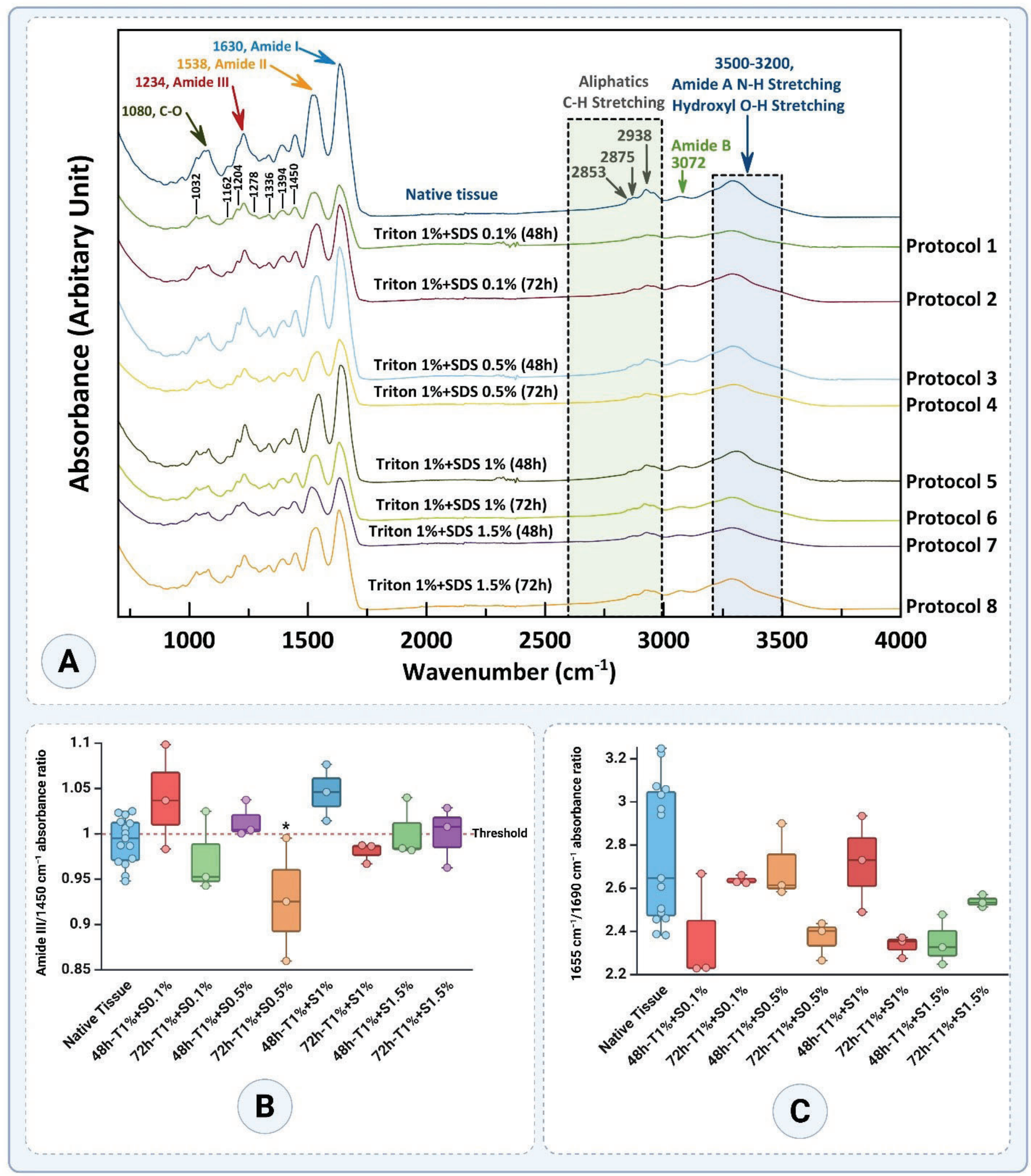

**Figure 6.** ATR-FTIR analysis of native and decellularized porcine uterine tissues following Triton™ X-100 and SDS treatments. (A) Representative normalized FTIR spectra highlighting characteristic ECM-related functional groups, including Amide I (~1,655 $cm^{-1}$, C=O stretching), Amide II (~1,538 $cm^{-1}$, N–H bending and C–N stretching), Amide III (~1,234 $cm^{-1}$, N–H bending and C–N stretching), aliphatic C–H stretching (2,938 and 2,875 $cm^{-1}$), and the broad –OH/–NH stretching region (3,500–3,200 $cm^{-1}$). (B) Amide III/1,450 $cm^{-1}$ absorbance ratios used as an internal indicator of collagen triple-helix preservation. (C) 1,655/1,690 $cm^{-1}$ absorbance ratios reflecting relative collagen secondary structure and crosslinking integrity across decellularization protocols. Statistical analysis was performed using one-way ANOVA followed by Dunnett's multiple comparisons test versus native tissue. Asterisks indicate significant differences from native tissue (*$p < 0.05$, **$p < 0.01$, ***$p < 0.001$, ****$p < 0.0001$). Groups without asterisks were not significantly different from native tissue. Data are presented as mean ± SD.

Overall, shorter treatment durations (48 h) generally preserved collagen integrity better than extended durations (72 h). Protocols with higher SDS concentrations, such as 1% and 1.5%, demonstrated effective preservation of collagen at shorter durations, with the T1% + S1% - 48h protocol achieving the highest ratio. However, prolonged exposure to these concentrations led to a decline in collagen integrity. These results emphasize the importance of balancing detergent concentration and exposure time to optimize decellularization while maintaining ECM structure.[41,49]

### *3.1.7. Raman spectroscopy analysis*

The Raman spectroscopic analysis revealed significant molecular changes in ECM components, particularly in proteins and glycosaminoglycans (GAGs), across 24, 48, and 72 h of decellularization treatments. The observed changes are characterized by shifts in Raman peaks, reflecting structural and biochemical alterations in collagen and GAGs. Figure 7A and Table 1 highlights key Raman peaks, including Amide I (1,660–1,680 $cm^{-1}$), Amide III (1,235–1,340 $cm^{-1}$), $CH_2/CH_3$ deformation (1,450 $cm^{-1}$), and skeletal stretches (930,950 $cm^{-1}$), which serve as molecular fingerprints of collagen and related ECM components.[50-52]

In Amide I Band (1,660–1,700 $cm^{-1}$) region, dominated by C=O stretching vibrations, reflects secondary structural motifs such as α-helices (~1,650 $cm^{-1}$) and β-sheets (~1,660–1,670 $cm^{-1}$). Native tissue showed strong signals, indicating a well-preserved structure. Among the treated groups, T1% + S0.1% - 48h showed an enhanced 1,660/1,620 ratio (1.68 ± 0.22), reflecting tight molecular packing and optimal β-sheet preservation (Figure 7B). However, prolonged exposure to detergents led to structural loosening, as evidenced by a decline in the ratio for T1% + S1.5% - 72 h (1.21 ± 0.15), indicative of molecular degradation.

The Amide III band (1,235–1,340 $cm^{-1}$), associated with N-H bending and C-N stretching vibrations, provided insights into the protein backbone dynamics. The native tissue's 1,240/1,270 ratio (1.10 ± 0.03) highlighted a balanced and intact structure (Figure 7C). In 48-h treatments, T1% + S1% - 48h (1.17 ± 0.07) demonstrated stress-induced conformational tightening, while T1% + S0.1% - 72h (0.86 ± 0.13) in the 72-h group reflected extensive backbone degradation.

The side-chain dynamics (1,450 $cm^{-1}$), corresponding to $CH_2/CH_3$ deformations, were evaluated through the 1,240/1,450 ratio. A decrease in this ratio in groups like T1% + S1.5% - 72h (0.68 ± 0.09) signaled weakened molecular interactions due to excessive chemical exposure (Figure 7D). In contrast, the 48-h treatment T1% + S0.5% - 48h retained near-native characteristics (0.86 ± 0.09), indicating effective preservation of side-chain stability.

The skeletal stretches (930–950 $cm^{-1}$), reflecting the backbone flexibility of collagen, were well-preserved in 48-h treatments like T1% + S0.5% - 48h. However, 72-h treatments, such as T1% + S1.5% - 72h, exhibited reductions in intensity, indicating significant backbone disruptions.

The sulfated GAGs (~1,063 $cm^{-1}$), attributed to S=O stretching vibrations, served as markers for the presence of GAGs. The 48-h treatment T1% + S0.5% - 48h retained strong signals, demonstrating effective preservation of sulfated GAGs, while these signals were markedly reduced in 72-h treatments like T1% + S1% - 72h, highlighting significant GAG depletion.

Comparisons across the different time points underscored the balance achieved in the 48-h treatments. While the 24-h treatments, such as T1% + S1% - 24h, exhibited tight conformational packing with an elevated 1,660/1,620 ratio (1.45 ± 0.12), these early-stage protocols began to show minor reductions in GAG-specific signals, indicating the onset of ECM disruption. The 48-h treatments, particularly T1% + S0.5% - 48h and T1% + S1% - 48h, maintained structural stability across multiple markers. These protocols showed near-native values for the 1,240/1,270 (1.22 ± 0.15) and 1,240/1,450 (0.86 ± 0.09) ratios, alongside effective GAG retention. By contrast, the 72-h treatments were characterized by extensive degradation, with the 1,320/1,450 ratio dropping significantly in T1% + S1% - 72h (0.58 ± 0.12) and GAG-specific peaks almost disappearing (Figure 7E).

Among the protocols studied, T1% + S0.5% - 48h emerged as the most effective. This treatment achieved a near-native balance in structural and biochemical markers, preserving collagen's secondary structure and maintaining GAG levels. While T1% + S1% - 48h also performed well, slight deviations in side-chain dynamics suggested that T1% + S0.5% - 48h offered superior preservation. These findings highlight the critical role of precise chemical exposure in optimizing decellularization outcomes.

### *3.1.8. Mechanical properties of decellularized tissues*

Tensile testing directly assesses the preservation of the

**Table 1. Raman shift assignments and structural implications for ECM components in decellularized tissue.**

| Raman shift ($cm^{-1}$) | Protein backbone/Functional group | Assignment | Secondary structure motif/ feature | Relevance in decellularization/ collagen studies | Ref. |
|---|---|---|---|---|---|
| 1,670–1,680 | Amide I | C=O stretching | β-turn/random β-space | Associated with structural reorganization or loss of secondary structure; often indicates stress-induced conformational changes. | 53-57 |
| 1,660–1,670 | | | β-sheet | Indicates tight molecular packing; sensitive to hydrogen bonding and chemical treatments. | |
| 1,650–1,655 | | | α-helix | Correlates with structural integrity; α-helical content is often reduced after harsh treatments like detergents. | |
| 1,330–1,340 | Amide III | N–H bending and C–N stretching | α-helix | Sensitive to protein backbone vibrations; alterations reflect denaturation or stress. | 57,58 |
| 1,235–1,250 | | | β-sheet | Indicative of β-sheet formation; associated with fibril stability. | |
| 930–950 | Skeletal stretch | C–C stretching | α-helix | Reflects backbone flexibility and stabilization. | |
| 1,600–1,610 | Aromatic side chains | C=C stretching (phenylalanine, tyrosine) | Aromatic ring vibration | Highlights changes in aromatic residues, indicative of chemical alterations. | 59 |
| 1,450 | $CH_2$/$CH_3$ deformation | $CH_2$/$CH_3$ bending | Side chain vibrations | Sensitive to hydration and side-chain interactions; decreased in degraded collagen. | 53 |
| 1,400–1,410 | $COO^-$ symmetric stretching | Carboxyl groups | Ionization and side-chain modifications | Reflects changes in acidic side chains during decellularization. | 60 |
| 1,080 | Skeletal stretch | C–O–C stretching | GAGs | Indicates preservation or removal of ECM components like GAGs. | 61,62 |
| 1,000 | Aromatic side chains | Phenylalanine ring breathing | Aromatic stability | Indicates changes in phenylalanine-rich environments, reflecting stress or treatment. | 63 |
| 860–870 | Hydroxyproline | C–C stretching in pyrrolidine ring | Collagen-specific vibrations | Strongly related to collagen stability; hydroxyproline is a marker of intact collagen. | 64 |
| 800–820 | Skeletal backbone | C–C bending | Backbone flexibility | Changes reflect decellularization impacts. | 65,66 |

Abbreviations: ECM: Extracellular matrix; GAGs: Glycosaminoglycans.

extracellular matrix (ECM), ensuring its structural and functional integrity post-decellularization. Damage to ECM components like collagen or elastin during processing can compromise its ability to support cell attachment, migration, and differentiation, reducing its effectiveness in tissue engineering.

The mechanical properties of native and decellularized tissues were evaluated based on stress-elongation behavior, ultimate tensile strength (UTS), elongation at rupture, and Young's modulus (Figure 8). The stress-elongation curves (Figure 8A) highlight the mechanical integrity of the native tissue, which demonstrated the highest stress values. Among the decellularized groups, the T1% + S0.1% - 48h treatment exhibited stress-strain behavior closest to the native tissue, suggesting minimal disruption to the tissue's mechanical properties. However, prolonged exposure (72 h) and increased detergent concentrations (0.5–1.5% SDS) led to a significant reduction in stress values, with T1% +

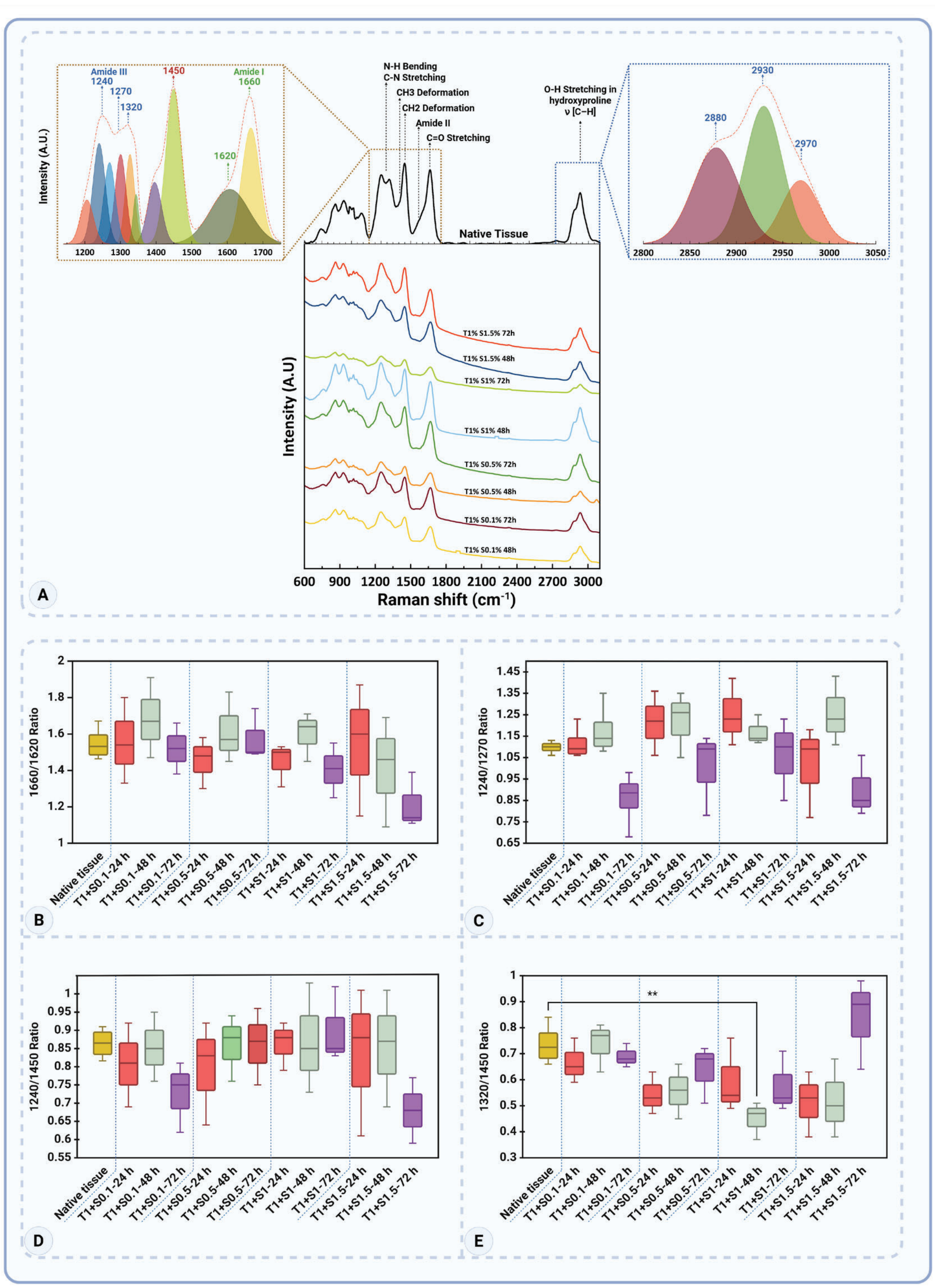

**Figure 7.** Comparative Raman Spectroscopic Analysis of ECM Components Post-Decellularization. (A) Characteristic Raman spectra for native and treated tissues across different protocols, highlighting key molecular regions such as Amide I, Amide III, $CH_2/CH_3$ deformations, and GAG-associated peaks. (B) The 1,660/1,620 ratio, reflecting molecular packing and β-sheet content. (C) The 1,240/1,270 ratio, indicative of protein backbone stability. (D) The 1,240/1,450 ratio, evaluating side-chain interactions and hydration. (E) The 1,320/1,450 ratio, measuring backbone-side-chain dynamics, comparing native tissue and treatments at 24, 48, and 72 h. Statistical analysis was performed using one-way ANOVA followed by Dunnett's multiple comparisons test versus native tissue. Asterisks indicate significant differences from native tissue (*$p < 0.05$, **$p < 0.01$, ***$p < 0.001$, ****$p < 0.0001$). Groups without asterisks were not significantly different from native tissue. Data are presented as mean ± SD.

S1% - 72h and T1% + S1.5% - 72h showing the weakest performance.

The UTS values (Figure 8B) followed a similar trend, where the native tissue exhibited the highest UTS (2.44 ± 0.05 MPa). T1% + S0.1% - 48h demonstrated a UTS comparable to the native tissue (2.34 ± 0.05 MPa, $p = 0.7094$), indicating that this protocol effectively preserved the tensile strength. However, the UTS declined significantly with increasing detergent concentrations and treatment durations. For instance, T1% + S0.5% -48h and T1% + S1.5% -72h exhibited markedly lower UTS values (1.39 ± 0.11 MPa and 1.19 ± 0.10 MPa, respectively, $p < 0.0001$). These findings emphasize that higher detergent concentrations and prolonged exposure compromise the tensile strength of decellularized tissues.

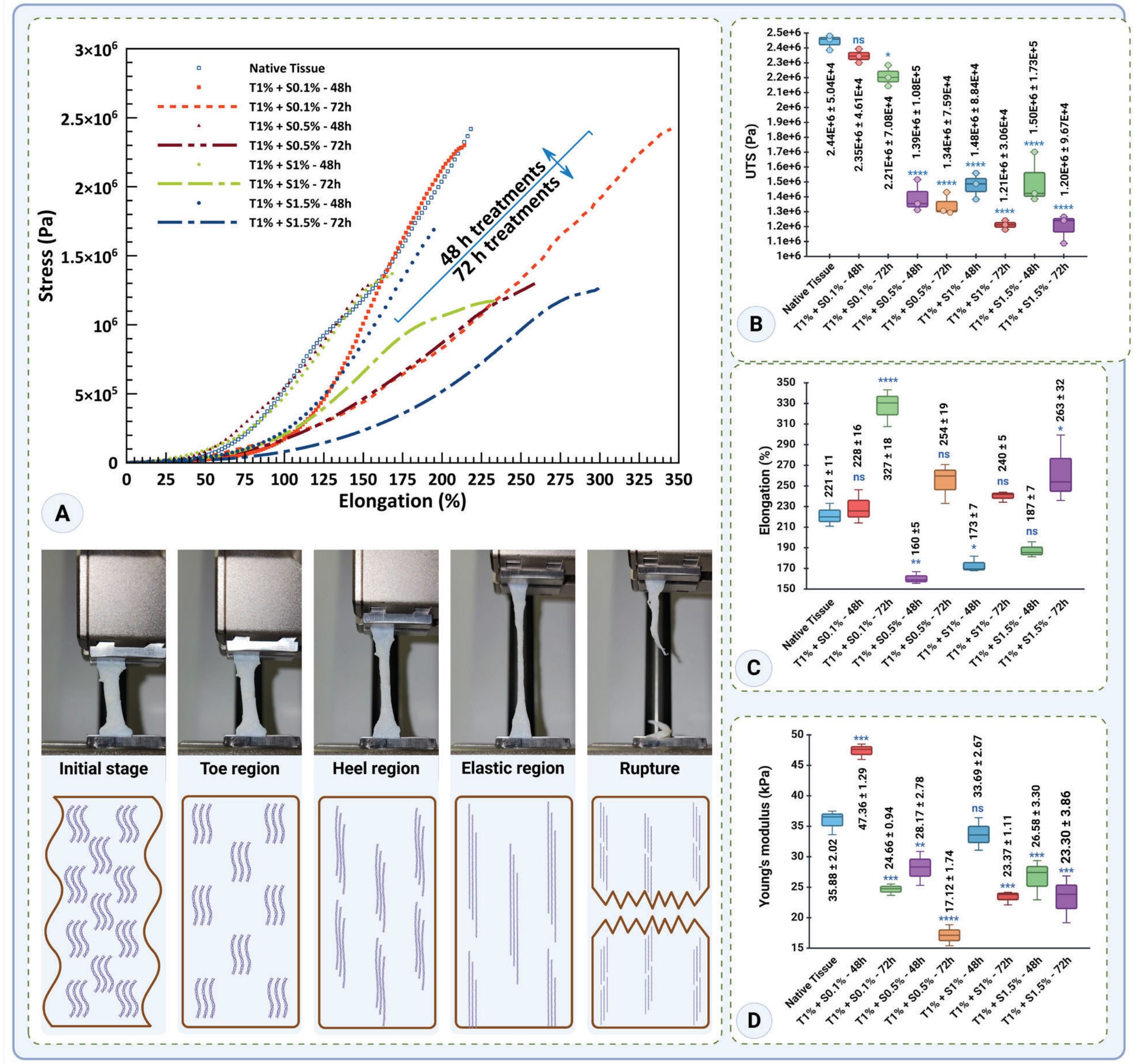

**Figure 8.** Mechanical characterization of native and decellularized tissues. (A) Stress-strain curves illustrating the mechanical behavior of native tissue and decellularized tissues treated with different protocols (T1% + SDS concentrations of 0.1%, 0.5%, 1%, and 1.5% for 48 and 72 h). The native tissue exhibited the highest stress values, while decellularized tissues showed reduced mechanical integrity, particularly at higher SDS concentrations and longer exposure times. (B) Ultimate tensile strength (UTS) values of native and decellularized tissues. The native tissue showed the highest UTS, while significant reductions were observed in treatments with higher SDS concentrations ($p < 0.0001$). (C) Percent elongation at rupture of the tissues. T1% + S0.1% - 72h demonstrated the highest elongation, indicating enhanced flexibility, while T1% + S0.5% - 48h and T1% + S1% - 48h had the lowest values. (D) Young's modulus (stiffness) of the tissues. The native tissue showed a modulus of 35.88 ± 2.02 kPa, with stiffness preserved in T1% + S0.1% - 48h but significantly reduced in treatments with prolonged exposure and higher SDS concentrations (p < 0.0001). Statistical significance is indicated as follows: ns = not significant, $p < 0.05$ (*), $p < 0.01$ (**), $p < 0.001$ (***), and *$p < 0.0001$ (****). Data are presented as mean ± SD.

Elongation at rupture, a measure of tissue flexibility, is presented in Figure 8C. The native tissue showed an elongation of 221.34 ± 11.13%. Among the decellularized groups, T1% + S0.1% - 72h exhibited the highest elongation (327.11 ± 18.07%, $p < 0.0001$), indicating enhanced flexibility compared to the native tissue. Conversely, treatments with higher SDS concentrations and shorter durations, such as T1% + S0.5% -48h and T1% + S1% -48h, significantly reduced elongation to 160.38 ± 5.81% and 173.01 ± 7.67%, respectively ($p < 0.01$). Notably, T1% + S1.5% -72h demonstrated a partial recovery in flexibility with an elongation of 263.04 ± 32.71% ($p < 0.05$).

Young's modulus, an indicator of tissue stiffness, is shown in Figure 8D. The native tissue exhibited a modulus of 35.88 ± 2.02 kPa. T1% + S0.1% -48h significantly increased stiffness to 47.35 ± 1.29 kPa ($p < 0.0001$), likely due to the retention of structural integrity. However, prolonged exposure (72 h) at 0.1% SDS reduced stiffness to 24.65 ± 0.94 kPa ($p < 0.0001$). The T1% + S0.5% - 72h treatment showed the lowest modulus (17.12 ± 1.73 kPa, $p < 0.0001$), indicating significant softening due to extended exposure. Intermediate stiffness values were observed for T1% + S1% -48h (33.68 ± 2.67 kPa) and T1% + S1.5% -48h (26.57 ± 3.30 kPa).

In summary, these results indicate that low detergent concentrations (0.1% SDS) and shorter exposure times (48 h) better preserve the mechanical properties of decellularized tissues. Conversely, higher concentrations and prolonged treatments significantly impair tensile strength, flexibility, and stiffness. These findings underscore the importance of optimizing decellularization protocols to maintain the mechanical integrity of the tissue for potential applications in tissue engineering.

#### *3.1.9. Selection of the optimal decellularization protocol for further analysis*

The evaluation of decellularization protocols revealed tissue-specific outcomes influenced by structural integrity. For tissues with less intact structures, such as the endometrium, and for more intact segments, such as the uterine myometrium, the T1% + S1% - 48h protocol demonstrated superior DNA removal, achieving levels below the critical threshold of 50 ng/mg. While this protocol showed slightly reduced protein preservation compared to T1% + S0.5% - 48h, its enhanced decellularization minimizes the risk of immunogenicity, making it more suitable for downstream applications. Conversely, T1% + S0.5% - 48h preserved more protein, which may be advantageous for maintaining ECM functionality. However, the higher DNA residue levels in this protocol limit its applicability due to potential immunogenic responses, especially in clinical and translational contexts.

Given these findings, the T1% + S1% - 48h protocol was selected for future studies to ensure a balance between structural preservation and immunological safety. The decellularized powdered dUECM generated using this protocol will be used for further biofabrications.

### 3.2. Alginate-thickened dUECM biomaterial ink formulation and rheological/printability characterization

Building upon the optimized dECM protocol, the second objective was to develop a printable biomaterial ink by incorporating uterine-derived dECM into alginate matrices. The rheological behavior, gelation kinetics, extrusion performance, and post-printing stability of the resulting alginate–dUECM biomaterial inks were evaluated.

#### *3.2.1. Gelation kinetics and rheological behavior of dUECM gel*

The steady-state rheological analysis of dUECM hydrogels at 0.5%, 1%, and 1.5% concentrations was conducted at 4 °C to evaluate their viscosity and stress behavior under varying shear rates (Figure 9A). The results revealed that increasing the concentration of dUECM led to higher viscosity, with the 1.5% dUECM hydrogel exhibiting the highest viscosity, followed by 1% and 0.5%. This trend indicates enhanced mechanical stability at higher concentrations. All samples displayed shear-thinning behavior, where viscosity decreased with increasing shear rate, a desirable characteristic for biomaterial ink applications as it facilitates smooth extrusion during extrusion-based 3D printing, while maintaining post-printing structural stability.

The gelation kinetics of the dUECM hydrogels were examined at 405 nm using turbidimetry (Figure 9B). The absorbance intensity increased over time, confirming gel formation. The 1.5% dUECM hydrogel exhibited the fastest gelation, initiating at approximately 6 min and completing by 20 min. The 1% dUECM hydrogel showed moderate gelation kinetics, with gelation beginning around 8 min and stabilizing by 22 min. In contrast, the 0.5% dUECM hydrogel demonstrated delayed gelation, initiating at approximately 10 min and stabilizing after 24 min. These results suggest that higher dUECM concentrations accelerate gelation, resulting in faster structural stabilization.

The microstructural organization of the hydrogels was further investigated using SEM imaging (Figure 9C–E). The pore size analysis revealed significant differences across the three concentrations. The 0.5% dUECM hydrogel had the largest mean pore size of 17.73 ± 4.48 μm (range: 10.09–25.91 μm), with a loose and sparsely connected network. The 1% dUECM hydrogel displayed a mean pore size of 13.39 ± 4.10 μm (range: 7.54–21.42 μm) and a more organized structure, characterized by visible crosslinked collagen fibers. The 1.5% dUECM hydrogel exhibited the smallest mean pore size of 11.92 ± 2.59 μm (range: 6.69–15.28 μm), with a compact and highly uniform structure. These findings highlight the influence of dUECM concentration on microstructural organization, where higher concentrations produce denser and more uniform networks.

The temperature-dependent gelation behavior of the 1.5% dUECM hydrogel was evaluated using G′ and G″ moduli to determine the impact of temperature on gelation kinetics (Figure $9F_{1\text{-}3}$). At 10 °C, gelation was initiated at 891 s, demonstrating a slow and controlled process suitable for storage and handling. Increasing the temperature to 15 °C accelerated gelation, with the onset occurring at 196 s. Further temperature increases to 20 °C, 25 °C, and 30 °C significantly reduced gelation times to 64, 39, and 27 s, respectively. These results confirm the thermoresponsive nature of dUECM hydrogels and underscore the critical importance of maintaining storage and handling conditions below 10 °C to prevent premature gelation.

In summary, the steady-state rheology results demonstrate enhanced mechanical stability with increasing dUECM concentration, while the gelation kinetics indicate faster gelation at higher concentrations. SEM imaging revealed that the microstructure becomes denser and more uniform with increasing concentration, and the temperature-dependent analysis highlighted the critical role of temperature in modulating gelation time. Together, these findings provide valuable insights into the design and application of dUECM hydrogels for various biomedical and biofabrication needs.

#### *3.2.2. Printability, swelling, degradation, and mechanical properties of extrusion 3D-printed constructs*

The printability of the 3D-printed constructs was assessed using the printability factor, where values closer to 1 indicate higher accuracy in reproducing the designed strand size. Because extrusion pressure and printing speed were adjusted for each formulation to achieve stable filament deposition, print fidelity metrics represent extrusion performance under formulation-specific printing conditions rather than intrinsic rheological properties alone. Six groups of hydrogels were analyzed: Alg 2% + 0.5, 1, and 1.5% dUECM and Alg 3% combined with the same dUECM concentrations. To ensure the hydrogels' effectiveness, their swelling, mechanical properties (Young's modulus), and degradation behavior were studied. Controlled swelling is vital for maintaining structural integrity, dimensional stability, and nutrient exchange, while appropriate mechanical properties ensure the hydrogels can support printed structures and mimic native tissue. Additionally, tailored degradation rates enable scaffold bioresorption, compatibility with tissue regeneration, and controlled release of bioactive molecules, making these hydrogels versatile for biomedical applications.

In the Alg 2% group, the printability factor for Alg 2% alone was 1.58 ± 0.46. When dUECM was incorporated, the printability factors were 1.60 ± 0.19 for 0.5% dUECM, 1.65 ± 0.30 for 1% dUECM, and 1.87 ± 0.25 for 1.5% dUECM (Figure $10A_{1\text{-}6}$). Statistical analysis showed no significant differences between Alg 2% and the formulations with 0.5% or 1% dUECM ($p > 0.05$). However, the difference for the formulation with 1.5% dUECM was statistically significant ($p < 0.0001$).

In the Alg 3% group, the printability factor for Alg 3% alone was 1.92 ± 0.21. When dUECM was added, the printability factors were 1.82 ± 0.18 for 0.5% dUECM, 1.20 ± 0.19 for 1% dUECM, and 1.56 ± 0.20 for 1.5% dUECM (Figure $10B_{1\text{-}6}$). Statistical analysis revealed no significant difference between Alg 3% and the formulation with 0.5% dUECM ($p > 0.05$). However, significant differences were observed for the formulations with 1% and 1.5% dUECM ($p < 0.0001$ for both).

The results for the Alg 2% group show that as the dUECM concentration increases, the printing pressure and printing speed are significantly affected (Figure $10A_{6\text{–}7}$). The mean pressure required for printing increased from 10.33 ± 0.58 kPa for Alg 2% alone to 11.00 ± 1.73 kPa for Alg 2% + dUECM 0.5%, 33.33 ± 5.77 kPa for Alg 2% + dUECM 1%, and 36.67 ± 5.77 kPa for Alg 2% + dUECM 1.5%. The difference in pressure was statistically significant between Alg 2% and formulations containing 1% ($p = 0.0003$) and 1.5% dUECM ($p = 0.0001$). This trend indicates that incorporating dUECM alters extrusion behavior, requiring higher extrusion pressure to maintain continuous filament deposition.

The mean printing speed followed a similar trend, increasing from 14.33 ± 1.15 mm/s for Alg 2% alone to 12.67 ± 2.52 mm/s for Alg 2% + dUECM 0.5%, 39.67 ± 6.81 mm/s for Alg 2% + dUECM 1%, and 57.33 ± 4.62 mm/s for Alg 2% + dUECM 1.5%. Statistical analysis revealed significant differences in printing speed for formulations containing 1% ($p = 0.0002$) and 1.5% dUECM ($p < 0.0001$). The increase in speed likely reflects parameter adjustments required to maintain strand uniformity under higher extrusion pressures.

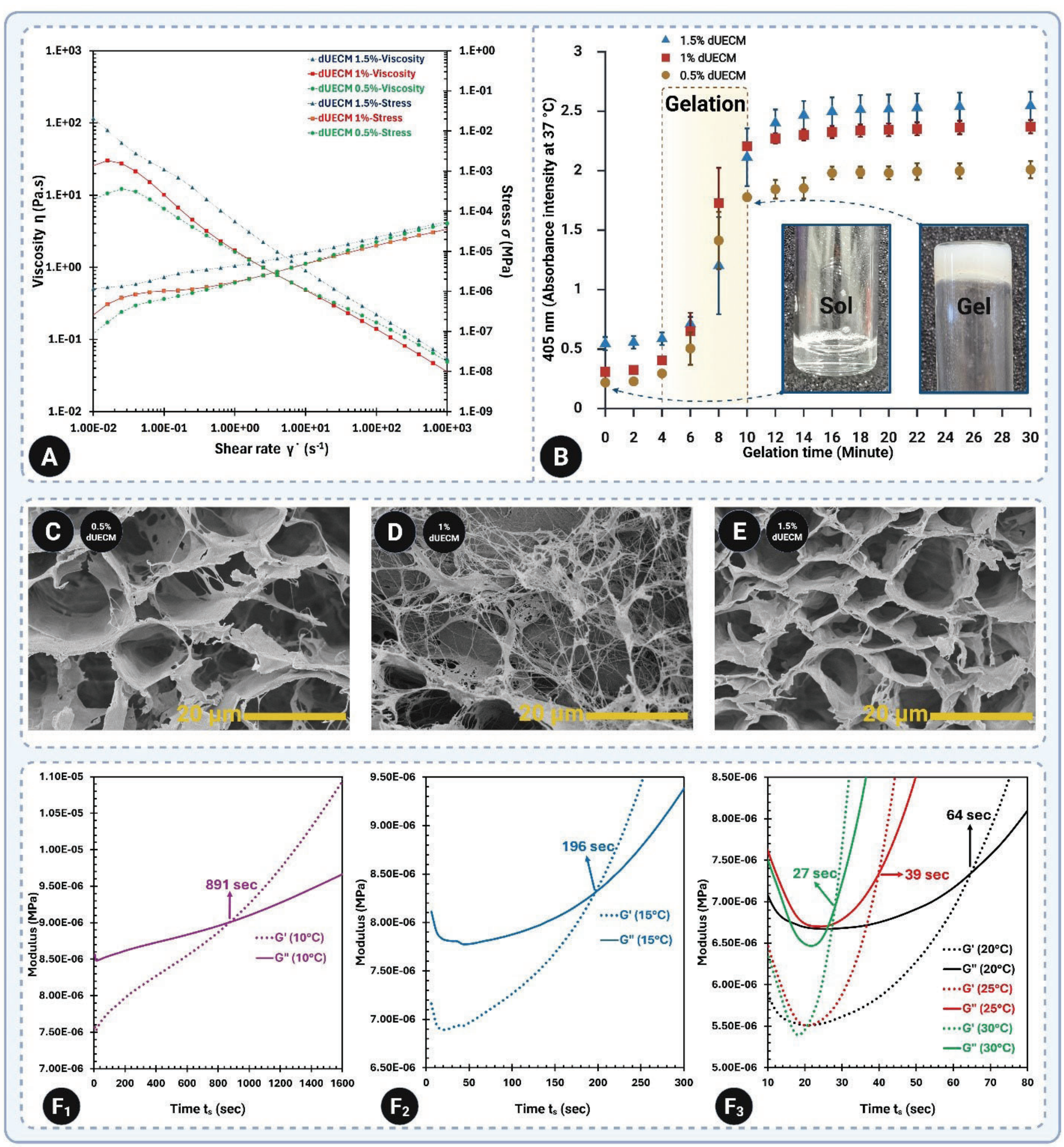

**Figure 9.** Gelation kinetics, rheological properties, and microstructural evaluation of dUECM hydrogels at different concentrations. (A) Steady-state rheological analysis at 4 °C showing viscosity and stress as functions of shear rate for 0.5%, 1%, and 1.5% dUECM hydrogels. All formulations exhibited shear-thinning behavior, with viscosity increasing as dUECM concentration increased. (B) Turbidimetric analysis at 405 nm showing concentration-dependent gelation progression over time. Inset images show representative sol and gel states of the hydrogels. (C–E) SEM micrographs showing the concentration-dependent microstructural organization of freeze-dried dUECM hydrogels. The 0.5% dUECM hydrogel showed a loose and sparsely connected network with larger apparent pores (C), while the 1% dUECM hydrogel displayed a more distinct fibrillar and interconnected ECM architecture with visible collagen-rich fibers (D). The 1.5% dUECM hydrogel exhibited a denser and more compact network with smaller and more uniform apparent pores (E). Apparent pore-size analysis was performed from SEM images acquired at the same magnification by measuring clearly identifiable open pores from representative regions, excluding image borders, collapsed areas, damaged regions, and preparation-related artifacts. The 1% dUECM image was selected to highlight the visible fibrillar, collagen-rich ECM network observed at this concentration. ($F_{1-3}$) Temperature-dependent changes in G′ and G″ of 1.5% dUECM hydrogels, showing faster gelation onset with increasing temperature. Data demonstrate that dUECM hydrogels exhibit concentration- and temperature-dependent gelation behavior, rheological properties, and microstructural organization.

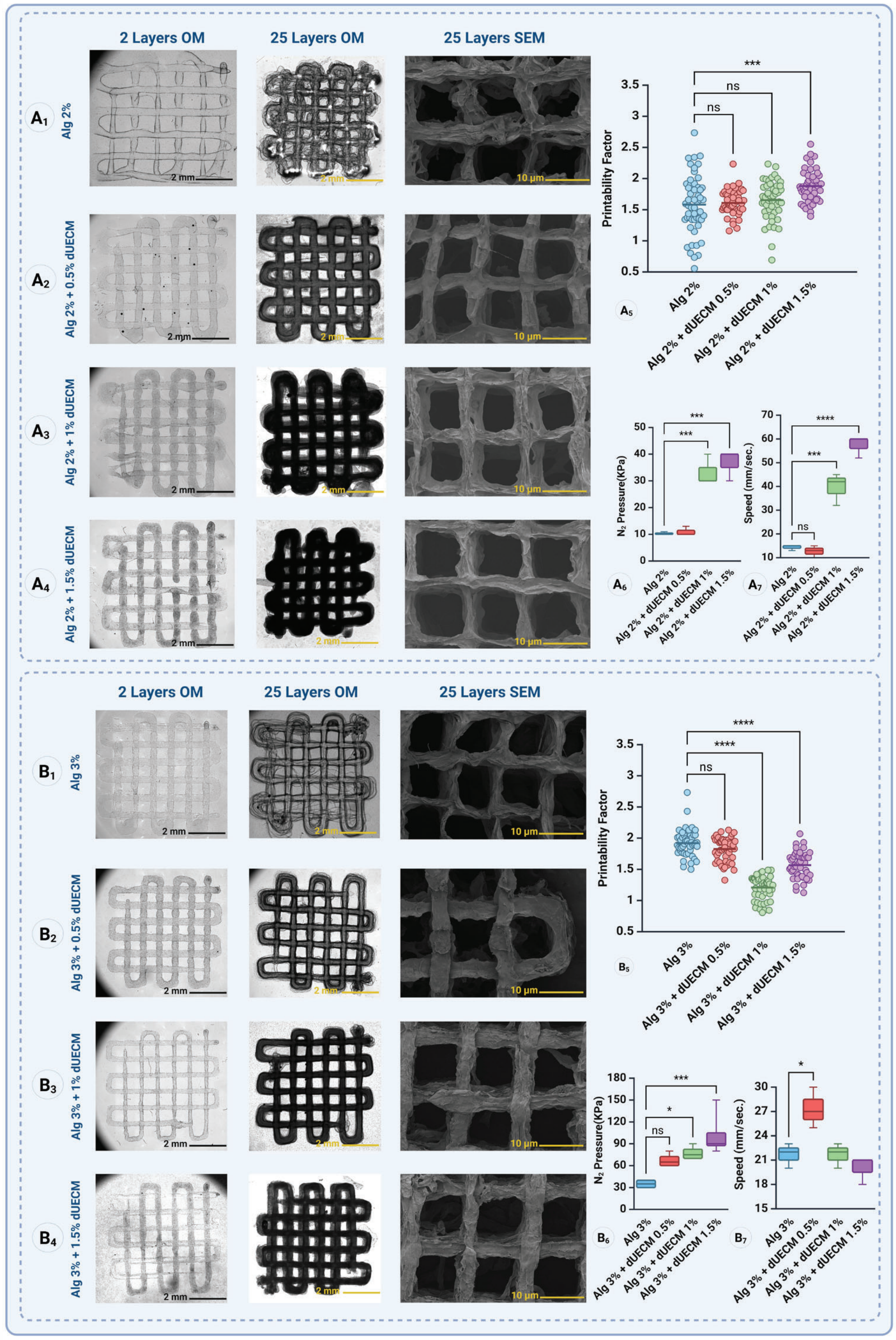

**Figure 10.** Printability and structural evaluation of 3D-printed constructs. Panels $A_1$–$A_4$ and $B_1$–$B_4$ show the optical microscopy (OM) images of 2-layer and 25-layer constructs and scanning electron microscopy (SEM) images of 25-layer constructs for the Alg 2% group ($A_1$–$A_4$) and Alg 3% group ($B_1$–$B_4$) with varying concentrations of dUECM (0.5%, 1%, and 1.5%). The images demonstrate the structural fidelity and strand morphology influenced by dUECM incorporation. Panels $A_5$ and $B_5$ depict the printability factors, with values closer to 1 indicating higher accuracy in reproducing the designed strand size. Panels $A_6$ and $B_6$ present the measured strand dimensions, with statistical comparisons showing the effects of dUECM addition on strand size. Statistical significance is indicated as follows: ns = not significant, $p < 0.05$ (*), $p < 0.01$ (**), $p < 0.001$ (***), and $p < 0.0001$ (****). Data are presented as mean ± SD.

In the Alg 3% group, a different trend was observed, where printing pressure increased consistently with higher dUECM concentrations, but printing speed showed a more variable behavior (Figure $10B_{6–7}$). The mean printing pressure increased from 35.00 ± 5.77 kPa for Alg 3% alone to 67.50 ± 9.57 kPa for Alg 3% + dUECM 0.5%, 77.50 ± 9.57 kPa for Alg 3% + dUECM 1%, and 102.50 ± 32.02 kPa for Alg 3% + dUECM 1.5%. The pressure differences were significant for formulations containing 1% ($p = 0.0134$) and 1.5% dUECM ($p = 0.0004$). The increase in pressure highlights the influence of higher dUECM concentrations on the biomaterial ink's extrusion properties.

The mean printing speed was 21.67 ± 1.53 mm/s for Alg 3% alone, which increased to 27.33 ± 2.52 mm/s for Alg 3% + dUECM 0.5%, and then decreased slightly to 21.67 ± 1.53 mm/s for Alg 3% + dUECM 1% and 20.00 ± 1.73 mm/s for Alg 3% + dUECM 1.5%. Statistical analysis showed a significant difference in speed only for the formulation with 0.5% dUECM ($p = 0.015$), with no significant differences for the 1% and 1.5% dUECM formulations ($p > 0.5$). The drop in speed for higher dUECM concentrations may reflect the need for finer control during printing under higher pressures.

The results indicate that incorporating dUECM at varying concentrations has a distinct impact on the printability factor across the two groups. Additionally, reduced gel translucency indicated the presence of bioactive dUECM molecules in the strands. These findings underscore the importance of further evaluating related properties, such as crosslinking behavior and swelling dynamics. Conducting swelling measurements over extended periods will provide deeper insights into how printability correlates with long-term structural stability and functional performance.

The swelling behavior and mechanical stability of the printed scaffolds were evaluated over 14 days in complete DMEM culture medium (Figure 11). Because hydrogel swelling equilibrium is typically reached within the first few hours, the extended 14-day swelling experiment was interpreted as a measure of long-term wet-mass stability and combined swelling–degradation behavior rather than short-term equilibrium swelling alone.

All scaffolds showed early water uptake within the first 24 h, consistent with hydration of the calcium-crosslinked alginate–dUECM network (Figure $11A_{1–2}$). In the 2% alginate-based groups, swelling generally decreased over time, suggesting gradual network relaxation, ionic exchange between calcium crosslinks and ions in the culture medium, and progressive scaffold degradation. Incorporation of dUECM reduced long-term swelling compared with alginate alone, particularly in the Alg 2% + 1% dUECM and Alg 2% + 1.5% dUECM groups, indicating improved wet-state stability.

A similar composition-dependent trend was observed in the 3% alginate-based scaffolds. The Alg 3% group showed the highest long-term swelling, whereas incorporation of dUECM limited excessive water uptake and produced more stable swelling profiles. Among these formulations, Alg 3% + 1.5% dUECM maintained a relatively stable swelling ratio over 14 days, suggesting that the combination of higher alginate concentration and dUECM reinforcement improved water retention stability without excessive swelling. Overall, these findings indicate that dUECM incorporation modulates scaffold hydration behavior and contributes to improved long-term wet-mass stability.

Mechanical testing further demonstrated that scaffold stiffness decreased over time in all groups, reflecting hydration-induced softening and structural remodeling during incubation (Figure $11B_{1-2}$). Pure alginate scaffolds showed the greatest reduction in Young's modulus, indicating limited mechanical stability under prolonged culture conditions. In contrast, dUECM-containing scaffolds retained stiffness more effectively, particularly at higher alginate and dUECM concentrations.

Among the 2% alginate-based scaffolds, incorporation of dUECM improved stiffness retention compared with alginate alone, with the Alg 2% + 1% dUECM group showing better mechanical stability over 14 days. In the 3% alginate-based formulations, the addition of dUECM further enhanced mechanical retention, with Alg 3% + 1% dUECM and Alg 3% + 1.5% dUECM maintaining the highest stiffness after prolonged incubation. These results suggest that dUECM contributes to the mechanical reinforcement of the alginate network, likely through matrix entanglement, ECM-derived structural components, and improved scaffold cohesion.

Together, the swelling and mechanical data demonstrate that dUECM incorporation improves the long-term stability of alginate-based printed scaffolds. The Alg 3% + 1.5% dUECM formulation provided the most favorable balance between controlled swelling behavior, stiffness retention, and structural integrity, supporting its selection as the optimized scaffold formulation for uterine tissue engineering applications (Figure $11B_3$).

The degradation behavior, illustrated in Figure $11C_1$ and $11C_2$, was assessed based on mass retention, where 100% represents the initial mass immediately after printing and crosslinking (day 0).

For the Alg 2% group, the scaffolds retained 100 ± 13% of their initial mass at day 0. By 24 h, the mass decreased to 78 ± 12% and further dropped to 53 ± 13% at day 3, 44 ± 14% at day 7, and 30 ± 12% by day 14. The incorporation of dUECM improved degradation resistance. Alg 2% + 0.5% dUECM scaffolds retained 100 ± 4% of their mass at day 0, which decreased to 76 ± 11% at 24 h, 66 ± 7% at day 3, 61 ± 5% at day 7, and 59 ± 8% by day 14. Alg 2% + 1% dUECM scaffolds retained 100 ± 12% initially, dropping to 77 ± 12% at 24 h, 69 ± 10% at day 3, 59 ± 9% at day 7, and stabilizing at 66 ± 12% by day 14. Finally, the Alg 2% + 1.5% dUECM scaffolds exhibited excellent degradation resistance compared to other Alg 2% groups. The scaffolds retained 100 ± 23% of their initial mass at day 0. By 24 h, the retention decreased slightly to 98 ± 21%, followed by 96 ± 22% on day 3, maintaining 96 ± 13% on day 7, and reducing to 94 ± 18% by day 14.

For the Alg 3% group, the scaffolds retained 100 ± 5% of their initial mass at day 0, with a degradation profile of 92 ± 6% at 24 h, 80 ± 10% at day 3, 78 ± 8% at day 7, and 66 ± 12% by day 14. Adding dUECM significantly enhanced stability. Alg 3% + 0.5% dUECM scaffolds retained 100 ± 13% of their mass at day 0, 95 ± 14% at 24 h, 92 ± 10% at day 3, 89 ± 11% at day 7, and 79 ± 14% by day 14. Alg 3% + 1% dUECM scaffolds started with 100 ± 14%, retained 95 ± 14% at 24 h, 94 ± 10% at day 3, 93 ± 11% at day 7, and maintained 91 ± 14% at day 14.

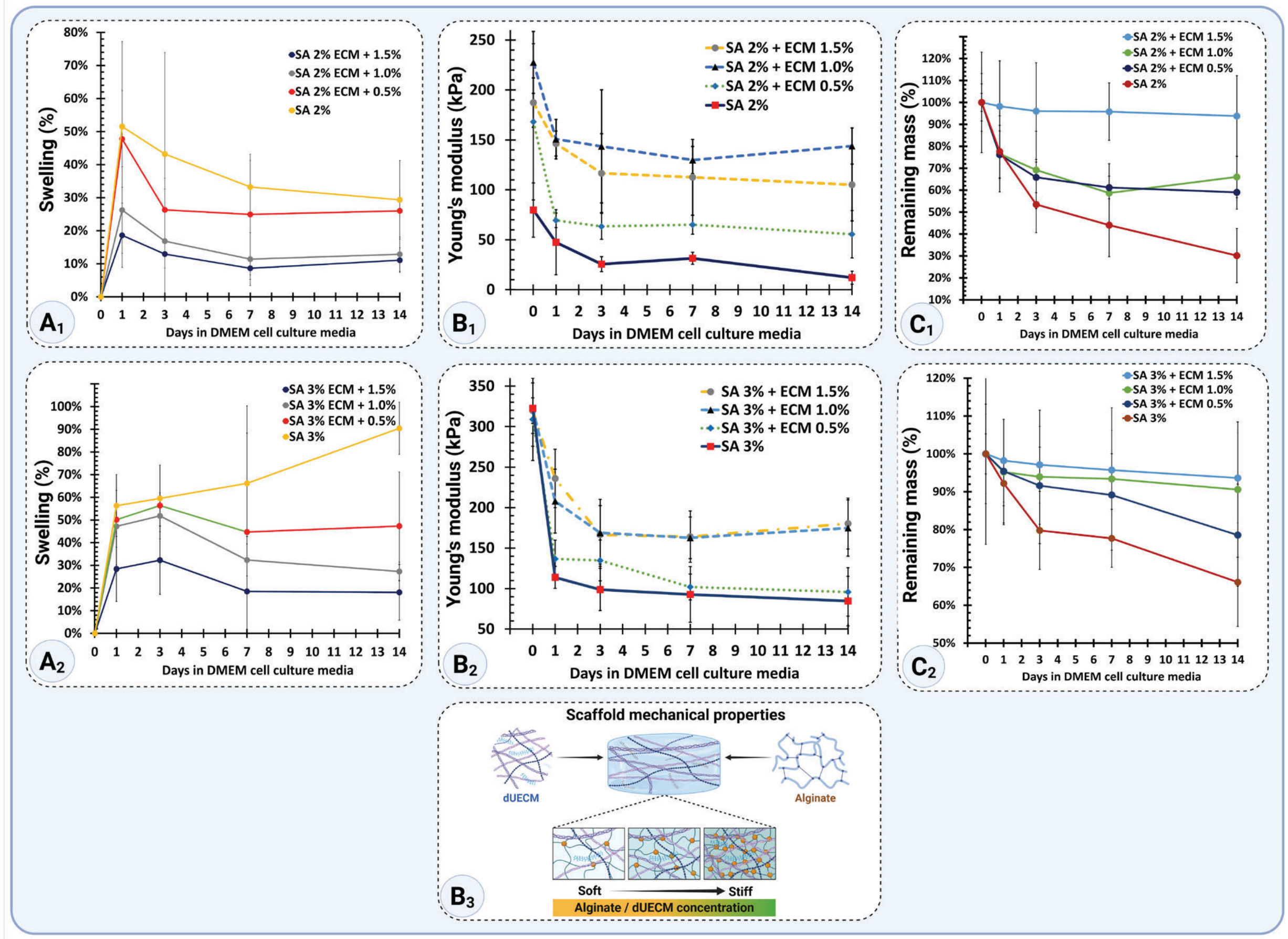

**Figure 11.** Swelling, mechanical, and degradation properties of alginate-based scaffolds supplemented with varying concentrations of decellularized extracellular matrix (dUECM). Panels $A_1$ and $A_2$: Swelling percentages of 2% alginate groups (SA 2%:dUECM 0.5, 1 and 1.5%) and 3% alginate groups (SA 2%:dUECM 0.5, 1 and 1.5%) scaffolds, respectively, over 14 days in DMEM culture media. Panels $B_1$ and $B_2$: Young's modulus of 2% and 3% alginate scaffolds, respectively, demonstrating the effect of dUECM concentrations (0.5%, 1.0%, and 1.5%) on scaffold mechanical stability over 14 days. Panel $B_3$: Schematic representation of scaffold mechanical properties showing the structural interplay between alginate and dUECM, where higher dUECM concentrations increase scaffold stiffness and stability. Panels $C_1$ and $C_2$: Degradation profiles based on the percentage of remaining mass of scaffolds, highlighting the enhanced stability of scaffolds with increasing dUECM concentrations.

Alg 3% + 1.5% dUECM scaffolds showed the highest degradation resistance among all groups. The scaffolds retained 100 ± 23% of their initial mass at day 0. By 24 h, the mass retention slightly decreased to 98 ± 21%, followed by 96 ± 22% at day 3, maintaining 96 ± 13% at day 7, and reducing to 94 ± 18% by day 14.

Taken together, these results indicate that alginate-thickened dUECM biomaterial inks possess formulation-dependent extrusion behavior, stable print fidelity under formulation-specific printing conditions, and improved resistance to degradation, particularly at 1.5% dUECM, supporting their use in multilayer extrusion-based 3D printing and justifying further investigation of their biological performance.

### 3.3. Biological performance of alginate–dUECM

The final objective of this study was to assess the biological performance of alginate–dUECM constructs fabricated using the optimized biomaterial ink formulations. Cellular viability, morphology, and tissue-relevant microstructural features were evaluated to determine their suitability for uterine tissue engineering applications.

#### *3.3.1. Cell viability, proliferation, and phenotypic assessment of hTERT-HM cells on alginate–dUECM hydrogel*

The results of the MTT assay reveal significant differences in relative hTERT-HM cell survival rates across various hydrogel compositions compared to the control groups. The positive control (plate without gel) demonstrated the highest cell survival, serving as the baseline at 100%. The negative controls (2% Alg and 3% Alg) exhibited significantly lower cell survival rates, serving as benchmarks for comparison with other experimental groups.

After 24 h, the positive control was set at 100% cell viability. The negative controls, 2% Alg and 3% Alg, showed reduced viabilities of 71.55 ± 5.97% and 68.66 ± 2.01%, respectively ($p = 0.026$ and $p = 0.0056$, respectively). Additional dUECM groups demonstrated intermediate viability, confirming the dose-dependent bioactivity of dUECM (Figure 12A).

By 72 h, 3% Alg + 1.5% dUECM maintained its lead at 102.53 ± 3.10%, exceeding both alginate-only controls ($p < 0.0001$). Among experimental groups, 3% Alg + 1.5% dUECM again led with 102.53 ± 3.10%, significantly higher than 2% and 3% alginate negative controls ($p < 0.0001$ and $p = 0.0005$, respectively). 3% Alg + 1% dUECM also performed well (98.25 ± 1.05%), confirming sustained bioactivity over time (Figure 12B). Other promising groups included 3% Alg + 1% dUECM, which achieved 98.25 ± 1.05%, indicating a stable and bioactive composition over time (Figure 12B).

At 120 and 168 h, the proliferative advantage of 3% Alg + 1.5% dUECM became increasingly pronounced (144.13 ± 6.39% and 258.14 ± 12.83%, respectively; $p < 0.001$). The group 2% Alg + 1.5% dUECM also maintained its high performance with a viability of 115.29 ± 2.76%, demonstrating the impact of dUECM in promoting cell proliferation (Figure 12C). and $p = 0.0003$, respectively), demonstrating a clear concentration and time-dependent increase in cell proliferation driven by dUECM bioactivity (Figure 12C and 12D). Full group-level data are reported in the figures.

Overall, the consistent superiority of 3% Alg + 1.5% dUECM across all days underscores its promise as a robust hydrogel composition for uterine tissue engineering. The Live/Dead assay further corroborates these findings, with 3% Alg + 1.5% dUECM displaying the highest proportion of live cells and minimal cell death throughout the study period.

The live/dead assay results, shown in Figure 12E and 12F, illustrate the hTERT-HM cell viability over 24, 72, 120, and 168 h for two groups: 3% Alg as the control (Figure 12E) and 3% Alg + 1.5% dUECM, the selected optimal composition based on MTT assay results (Figure 12F).

At 24 h, we observed a sparse distribution of live (green) cells, with significant red fluorescence, suggesting limited cell attachment and viability. At the 72-h mark, there was a small increase in cell viability, noticeable via the greater number of green cells. However, the red fluorescence of dead cells remains prominent, indicating persistent cell death. At 120 h, live cells appear more densely populated, but the overall viability remains moderate. After 168 h, there is a noticeable increase in green fluorescence, suggesting an improved environment for cell survival, but red fluorescence is still prevalent, reflecting limited bioactivity of 3% Alg alone.

In contrast, the 3% Alg + 1.5% dUECM group demonstrates a significant improvement in cell viability across all time points. After 24 h, green fluorescence is markedly more prominent than the control, indicating enhanced initial cell attachment and viability due to the bioactive properties of dUECM. By 72 h, cells exhibit increased proliferation and spreading, with minimal red fluorescence. By 120 h, the density of green cells significantly rises, showing active proliferation and reduced cell death.

After 168 h, the 3% Alg + 1.5% dUECM composition supports robust cell growth, with dense green fluorescence and minimal red fluorescence, reflecting its superior capacity to maintain a favorable microenvironment for hTERT-HM cells.

These results confirm the enhanced bioactivity of 3% Alg + 1.5% dUECM, which not only supports higher cell viability but also reduces cell death over time compared to 3% Alg alone. This highlights its potential as a promising hydrogel for uterine tissue engineering applications.

Immunohistochemical staining for α-smooth muscle actin (α-SMA) further supported these observations (**Figure 12E** and **12F**). hTERT-HM cells cultured on 3% Alg + 1.5% dUECM displayed pronounced α-SMA organization, elongated spindle-like morphology, and increased cytoskeletal alignment over time, particularly at days 5 and 7. In contrast, cells on 3% Alg alone showed weaker α-SMA signal intensity, less organized cytoskeletal structure, and reduced cell spreading. These findings indicate that incorporation of dUECM not only enhances cell viability and proliferation but also promotes maintenance of smooth-muscle–relevant cytoskeletal organization.

### *3.3.2. Physicochemical and microstructural characterization of cast/freeze-dried Alg–dUECM scaffolds*

To further evaluate the versatility of the optimized hydrogel formulation beyond extrusion-based printing, the selected

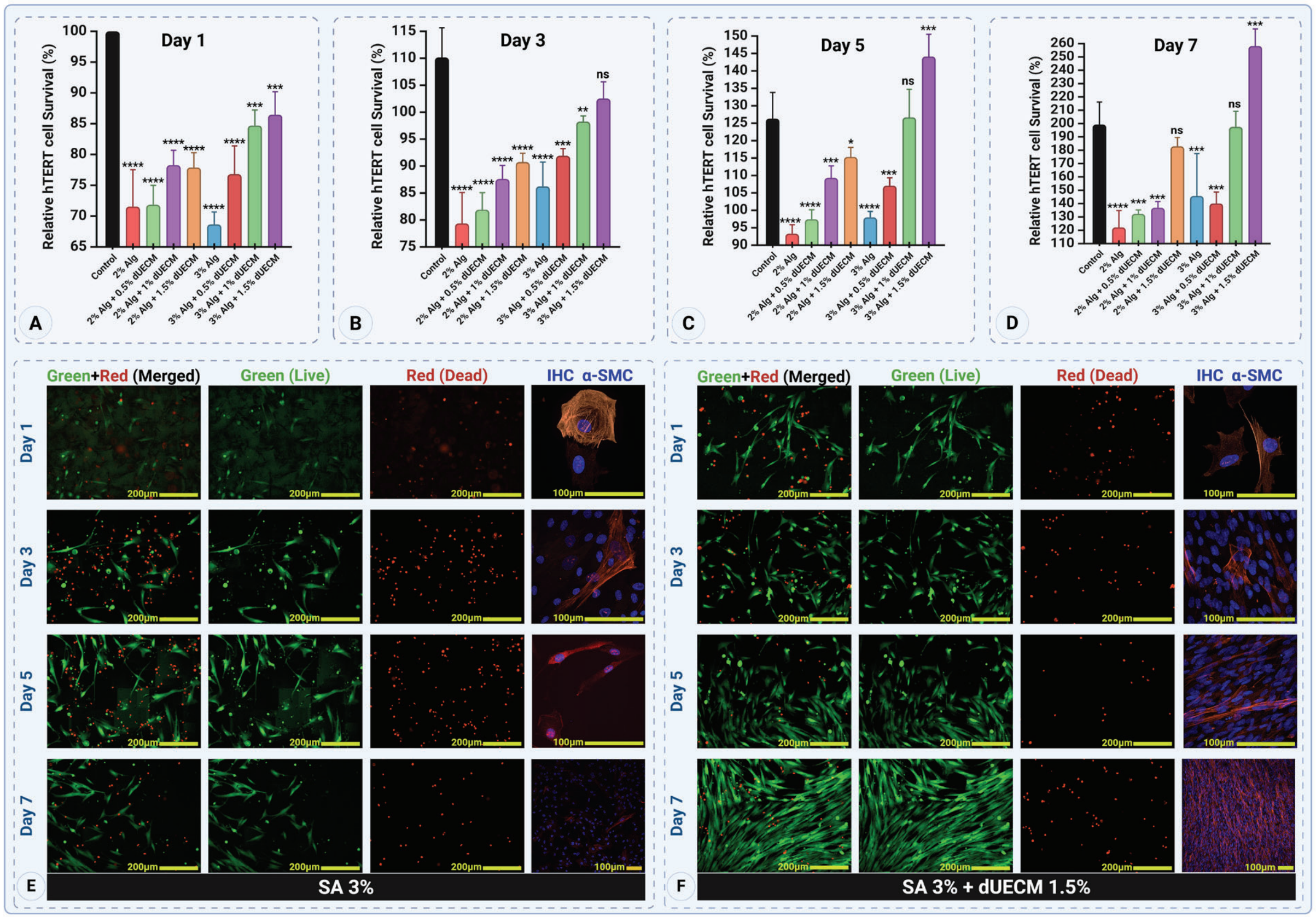

**Figure 12.** Cell viability, proliferation, and cytoskeletal organization of hTERT-HM cells cultured on alginate and alginate–dUECM hydrogels. (A–D) Relative cell survival (%) of hTERT-HM cells cultured on different hydrogel formulations for day 1 (A), day 3 (B), day 5 (C), and day 7 (D), as assessed by the MTT assay. Data are normalized to the control and presented as mean ± SD ($n = 3$). (E,F) Representative Live/Dead fluorescence images of hTERT-HM cells cultured on 3% SA (E) and 3% SA + 1.5% dUECM hydrogels (F) at days 1, 3, 5, and 7. Live cells are shown in green (calcein-AM) and dead cells in red (propidium iodide). Immunohistochemical (IHC) staining for α-smooth muscle actin (α-SMA, red) with DAPI-counterstained nuclei (blue) illustrates cytoskeletal organization and cellular morphology. Compared to alginate-only scaffolds, the 3% SA + 1.5% dUECM hydrogels supported enhanced cell attachment, elongated spindle-like morphology, increased cell density, and more pronounced α-SMA organization over time, particularly at days 5 and 7. Scale bars: 200 μm (Live/Dead); 100 μm (α-SMA IHC). Scale bar: 200 μm (for Green+Red, Green, and Red); 100 μm (for IHC α-SMC). Statistical significance is indicated as follows: ns = not significant, $p < 0.05$ (*), $p < 0.01$ (**), $p < 0.001$ (***), and $p < 0.0001$ (****). Data are presented as mean ± SD.

Alg 3% + 1.5% dUECM hydrogel was also prepared using a casting and freeze-drying approach. This analysis was performed to assess whether the formulation could function as an alternative membrane- or mesh-like scaffold format for uterine tissue engineering. Crosslinked (CL) and non-crosslinked (NCL) scaffolds were compared to determine how ionic stabilization affects thermal behavior, molecular interactions, hydration-related features, and scaffold microarchitecture (Figure 13).

Thermal analysis demonstrated that crosslinking improved the stability of the Alg–dUECM scaffold. Compared with the non-crosslinked scaffold, the crosslinked hydrogel showed more defined thermal transitions, suggesting stronger matrix stabilization and improved retention of bound water within the hydrogel network (Figure 13A). This hydration-related behavior is important for maintaining scaffold integrity and supporting a physiologically relevant microenvironment for cell attachment, nutrient transport, and tissue integration.

FTIR analysis further confirmed that crosslinking altered the molecular organization of the Alg–dUECM hydrogel (Figure 13B). The crosslinked scaffold exhibited more defined hydroxyl, amide, aliphatic C–H, and C–O stretching regions, indicating enhanced hydrogen bonding, preservation of dUECM-associated protein features, and interactions between alginate and ECM-derived components. These molecular features support the role of crosslinking in improving scaffold cohesion while maintaining bioactive functional groups associated with the dUECM component.

SEM analysis was then used to compare the microstructure of native uterine tissue and the freeze-dried Alg–dUECM scaffolds (Figure 13C–E). Native uterine tissue showed layer-specific structural organization, including the endometrium, myometrium, and serosal surface. The crosslinked scaffold exhibited a smoother and more continuous surface, together with an interconnected porous architecture, suggesting improved structural cohesion after crosslinking. In contrast, the non-crosslinked scaffold showed a less uniform morphology with more irregular pore formation and globular polymer-rich regions, indicating weaker matrix organization.

Porosity analysis supported these morphological observations (Figure 13F). Native uterine tissue displayed layer-dependent pore-size variation, reflecting the different structural roles of uterine tissue compartments. The freeze-dried Alg–dUECM scaffolds showed pore sizes within a range relevant for tissue-engineering applications, with crosslinking producing a more compact and structurally refined porous network compared with the non-crosslinked scaffold. This reduction in pore size after crosslinking is likely related to matrix tightening and ionic stabilization of alginate chains.

Overall, these findings indicate that crosslinking improves the physicochemical stability and microstructural organization of cast/freeze-dried Alg 3% + 1.5% dUECM scaffolds. Together with the printability and biological data, this supports the potential use of the optimized Alg–dUECM formulation in multiple scaffold formats, including printed constructs and cast/freeze-dried membrane-like scaffolds for uterine tissue engineering.

## 4. Discussion

### 4.1. Efficiency of one-step decellularization vs. traditional methods

Conventional uterine decellularization protocols often employ multiple sequential detergents and extended processing times. While these multistep approaches can effectively remove cellular material, they increase the risk of cumulative damage to the ECM due to prolonged chemical exposure and repeated handling. In contrast, our one-step protocol using 1% Triton™ X-100 combined with graded SDS concentrations (0.1, 0.5, 1, and 1.5%) enabled rapid and efficient decellularization while minimizing processing time and reagent use, thereby reducing opportunities for ECM degradation. In contrast, our protocol employs a one-step approach in which Triton™ X-100 and SDS are combined simultaneously in a single detergent bath, rather than applied sequentially in separate processing cycles as in conventional multi-step protocols. It is important to clarify that the term "one-step" refers to this simultaneous combination strategy, not to the total calendar duration of the process. The complete workflow encompasses: (i) an initial 24-h DPBS wash to remove residual blood; (ii) 48 h of simultaneous Triton™ X-100 + SDS decellularization; (iii) 72 h of post-decellularization washing in Milli-Q water to eliminate residual detergent; and (iv) approximately 48 h of freeze-drying. In contrast, conventional sequential protocols employ each detergent for 24–48 h in discrete steps with intervening washes, resulting in total active chemical exposure times that frequently exceed 7–10 days, not accounting for primary washing or freeze-drying steps. The optimal protocol in the present study was determined empirically by systematically varying both SDS concentration (0.1–1.5%) and treatment duration (48 vs. 72 h), while concurrently evaluating four key criteria: (1) residual DNA content (target <50 ng/mg dry weight), (2) GAG retention, (3) collagen structural integrity assessed by FTIR and Raman spectroscopy, and (4) tissue-level mechanical properties. The Triton™ 1% + SDS 1% for 48 h protocol achieved the most favorable balance across all four parameters and was therefore selected as optimal for downstream applications. To understand the basis of this improved balance between decellularization efficiency and ECM preservation, it is important to consider the complementary mechanisms of action of Triton™ X-100 and SDS. The combination of Triton™ X-100 and SDS in decellularization protocols provides a synergistic approach that enhances cell removal efficiency while preserving the integrity of the ECM.[67,68] Triton™ X-100, a nonionic surfactant, disrupts lipid-lipid and lipid-protein interactions in cell membranes, facilitating the solubilization of cellular components with minimal damage to the ECM. Conversely, SDS, an anionic surfactant, efficiently disrupts cell membranes and solubilizes nuclear material due to its strong ionic interactions.[29] When mixed, Triton™ X-100 and SDS do not neutralize each other, as Triton™ X-100 lacks ionic charges that would oppose SDS's anionic activity. Instead, their combination forms mixed micelles, which optimize the surface activity of the solution.[69] This interaction enables SDS to retain its cellular removal capacity while Triton™ X-100 moderates SDS's harsh effects on ECM components, such as glycosaminoglycans and collagens.

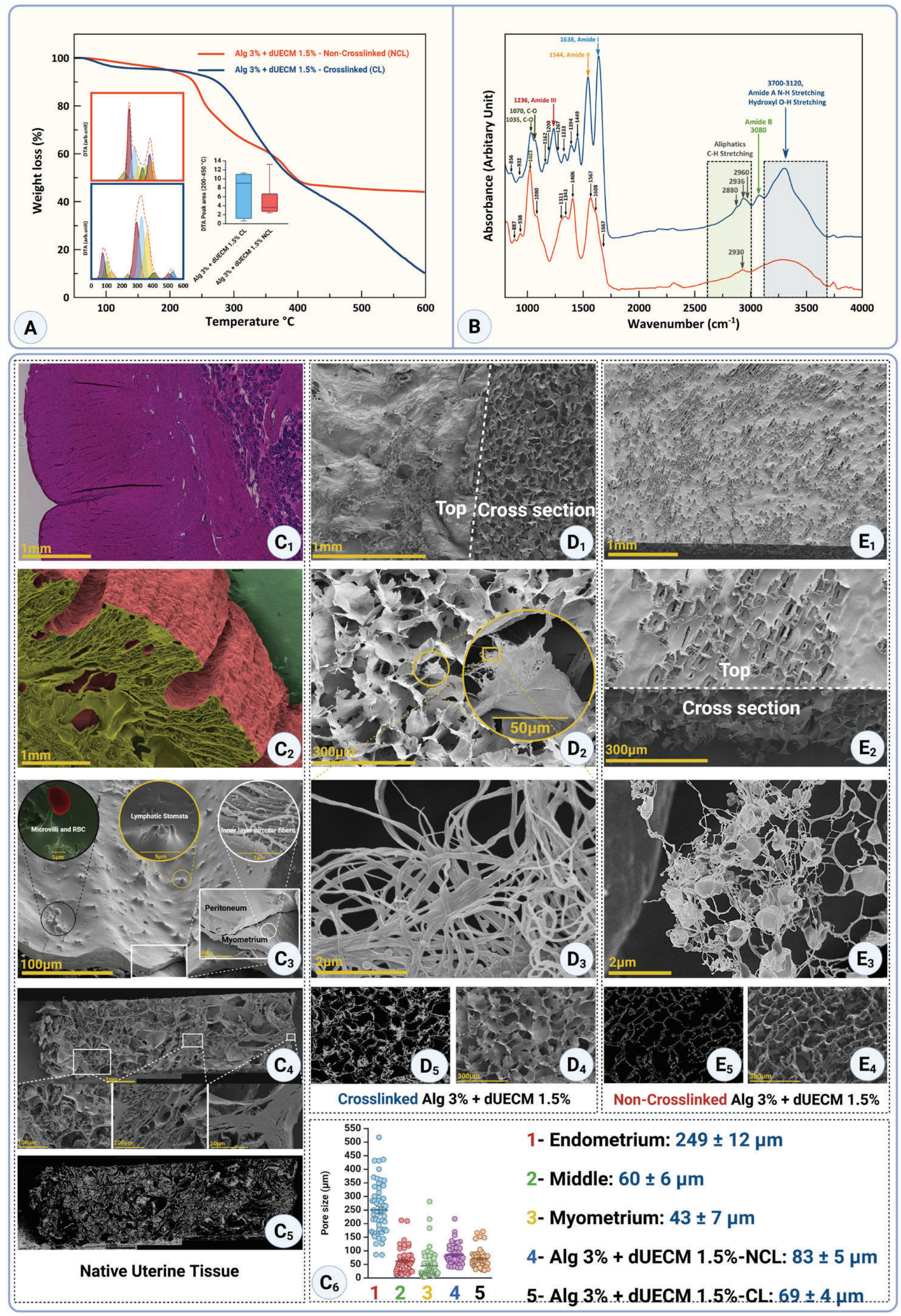

**Figure 13.** Thermal, chemical, and microstructural analysis of native uterine tissue and freeze-dried hydrogels composed of Alg 3% + dUECM 1.5%. (A) TGA and showing weight loss and thermal transitions of crosslinked (CL) and non-crosslinked (NCL) hydrogels. (B) FT-IR spectra comparing crosslinked and non-crosslinked hydrogels, emphasizing the preservation of key functional groups such as Amide I, II, and III, and the prominent hydroxyl group (3,700–3,120 cm$^{-1}$) in crosslinked hydrogels. ($C_1$–$C_6$) Microstructural characterization of native uterine tissue and hydrogels: ($C_1$) H&E-stained native uterine cross-section showing distinct layers; ($C_2$–$C_3$) SEM images of native uterine tissue depicting the serosa, lymphatic stomata, and aligned collagen fibers; ($C_4$–$C_5$) cross-sectional SEM and uterine porosity structure. ($D_1$–$D_5$) SEM images of crosslinked Alg 3% + dUECM 1.5%, highlighting an intact surface, uniform interconnected pores, and well-aligned collagen fibers. ($E_1$–$E_5$) SEM images of non-crosslinked Alg 3% + dUECM 1.5%, showing cylindrical porosities, globular alginate structures, and a disrupted surface morphology. ($C_6$) Quantitative analysis of pore size distribution in native uterine layers (endometrium, interlayer, and myometrium) and hydrogels (CL and NCL), confirming the mimicked porosity of crosslinked scaffolds for uterine tissue engineering applications.

Efficient cell removal was evidenced by DNA content reduced well below 50 ng/mg dry weight—a widely accepted threshold for minimizing immunogenicity.[7,22,70] For example, decellularized porcine uterine tissue processed with Triton 1% + SDS 1% for 48 h retained only ~51 ng/mg DNA, which marginally exceeds the commonly cited <50 ng/mg threshold. Three complementary considerations support the biological acceptability of this outcome: (i) gel electrophoresis confirmed complete absence of intact high-molecular-weight DNA in this group (Figure 3B), indicating that residual DNA is highly fragmented and non-chromatin-associated, substantially reducing its immunostimulatory potential; (ii) the 50 ng/mg criterion[22] is a practical benchmark rather than an absolute immunological threshold, and immunogenic consequence depends on DNA fragment size and chromatin integrity as much as on quantity; and (iii) the three protocols that achieved DNA strictly below 50 ng/mg (SDS 1% at 72 h, SDS 1.5% at both time points) all demonstrated significantly greater GAG depletion, spectroscopic evidence of collagen disorganization, and inferior mechanical properties. The T1% + SDS 1% at 48-h protocol, therefore, represents the best balance between decellularization efficacy and ECM preservation. This outcome is comparable to or better than more complex perfusion-based protocols. In Campo *et al.*'s study,[24] the residual DNA following the 0.1% SDS and 1% Triton X-100 protocol yielded 40.792 ± 16.462 ng/mg wet tissue, which when normalized to dry weight, substantially exceeds the 50 ng/mg threshold.

In Tiemann *et al.*'s study,[71] they also conducted perfusion with 0.5% SDS (8 h each) plus extensive washing to decellularize whole sheep uteri. Another key limitation of perfusion decellularization is that blockage of blood vessels and air trapping can disrupt uniform detergent flow, resulting in uneven decellularization and localized ECM damage, particularly in complex tissues.[72] One of the missing parts of the decellularization puzzle is the intrinsic dependence of double-stranded DNA (dsDNA) metrics on tissue source and normalization strategy. Marked differences in native DNA content across studies are expected because baseline cellularity varies substantially with tissue type and species, and because dsDNA values are often normalized to wet versus dry tissue mass; wet-weight reporting can substantially underestimate dsDNA relative to dry-weight normalization, thereby limiting direct cross-study comparisons to the commonly cited <50 ng/mg dry ECM benchmark.[73] Our simplified immersion-based decellularization approach achieved comparably high levels of DNA removal within a single processing cycle, an important advantage for scale-up and clinical translation where processing time and reagent burden are critical. Importantly, Triton-X100 1% + SDS 1% - 48 h protocol did not result in severe mechanical deterioration of the extracellular matrix. Prior uterine decellularization studies have shown that when mild detergent exposure is employed, cellular components can be effectively removed while the elastic modulus—primarily governed by collagen fibers—remains statistically unchanged, indicating preservation of the mechanically dominant collagen network despite detergent treatment.[74] Consistent with these observations, the decellularized uterine tissues in the present study retained robust mechanical performance, indicating that the one-step protocol avoided excessive disruption of the collagenous framework.

Notably, this outcome contrasts with reports in human uterine tissues subjected to prolonged, high-concentration SDS perfusion protocols (2% SDS for ~28 days followed by 5 days of PBS washing), where significant deterioration of mechanical properties was observed.[75] Together, these findings underscore that detergent concentration, exposure duration, and processing strategy are key determinants of mechanical preservation, highlighting the importance of minimizing overexposure to maintain translationally relevant uterine ECM mechanics.

### 4.2. ECM retention and SDS residue

A central challenge in decellularization is striking a balance between aggressive cell removal and the preservation of ECM components. We observed that detergent concentration and exposure time had a pronounced effect on collagen integrity and the retention of glycosaminoglycans (GAGs).

Higher SDS concentrations (1.5%) and longer treatments (72 h) yielded nearly complete DNA clearance but at the cost of significant GAG depletion and subtle collagen damage. Specifically, native uterine tissue contained ~54 μg GAG per mg, whereas the harshest protocol (Triton 1% + SDS 1.5%, 72 h) retained only ~30 μg/mg. This outcome is consistent with reports that SDS can disrupt the ECM by extracting GAGs and compromising collagen network integrity.[76]

In addition to direct ECM extraction, residual SDS retention emerged as an important mechanistic contributor to ECM compromise under harsher decellularization conditions. With the MBAS assay, we quantified residual SDS and observed a clear concentration-dependent increase across protocols, with low SDS conditions (0.1–0.5%) retaining minimal detergent (6–19 μg/g dry tissue) and higher concentrations showing substantially elevated residues, particularly at 1.5% SDS for 72 h (84.8 μg/g). These data indicate that increasing SDS concentration and exposure duration enhances detergent association with the ECM, likely through ionic interactions with collagen and proteoglycans. Importantly, the conditions exhibiting the highest SDS retention coincided with the greatest GAG loss and collagen disruption, suggesting that both direct detergent extraction and residual SDS association contribute to ECM degradation under aggressive protocols.[36]

The effect of SDS concentration and exposure duration on collagen molecular integrity was further elucidated using FTIR and Raman spectroscopy. In native collagen, the Amide I band (≈1,600–1,700 $cm^{-1}$) typically exhibits sub-peaks around 1,655 and 1,690 $cm^{-1}$, reflecting the ordered triple-helical structure of collagen.[41,77] Consistent with FTIR evidence that the 1,655/1,690 (Amide I) ratio tracks collagen supramolecular order and crosslink-associated organization, the SDS-dependent reduction in our 1,655/1,690 ratio suggests partial loss of collagen packing/ordering under harsher conditions.

Raman spectroscopy was used as a complementary, structure-sensitive technique to mechanistically validate the SDS-dependent ECM alterations identified by FTIR. Raman bands associated with collagen and glycosaminoglycans (GAGs), including Amide I (≈1,660–1,680 $cm^{-1}$; C=O stretching), Amide III (≈1,235–1,340 $cm^{-1}$; N–H bending and C–N stretching), $CH_2$/$CH_3$ deformation (~1,450 $cm^{-1}$), and sulfated GAGs (~1,063 $cm^{-1}$; S=O stretching), were consistent with established assignments for collagen-rich tissues.[66,78]

Across protocols, Raman analysis demonstrated that SDS concentration and exposure duration primarily affected collagen molecular organization rather than complete collagen removal. Protocols employing moderate SDS concentrations at 48 h (T1% + S0.5% and T1% + S1%) preserved near-native Amide I/Amide III relationships and collagen-to-matrix ratios (normalized to the ~1,450 $cm^{-1}$ band), indicating maintenance of collagen backbone order and triple-helical structure. In contrast, prolonged exposure and high SDS concentration (1.5% for 72 h) resulted in attenuation and redistribution of Amide III features, consistent with disruption of collagen hydrogen bonding and supramolecular packing rather than wholesale collagen loss.[79,80]

Collectively, Raman findings corroborate FTIR results and demonstrate that ECM degradation under aggressive decellularization conditions is driven by molecular disorganization of collagen and GAG extraction, providing a mechanistic explanation for the concurrent decline in mechanical performance. These data identify T1% + S1% for 48 h as an optimal decellularization window that balances efficient cell removal with preservation of collagen molecular architecture and ECM functionality.

It is acknowledged that additional immunostaining for specific collagen subtypes (*e.g.*, collagen types I and III), fibronectin, and laminin would provide greater specificity in ECM component characterization. However, the current multi-modal panel, comprising H&E, DAPI, Masson's trichrome, Picrosirius Red under polarized light, Alcian Blue, DMMB-based GAG quantification, ATR-FTIR, and Raman spectroscopy, collectively provides comprehensive coverage of the major structural and biochemical ECM components across all decellularization conditions. The combination of quantitative biochemical assays and molecular spectroscopic techniques offers a level of ECM characterization rigor that is consistent with or exceeds the standards of comparable decellularization studies in the uterine tissue engineering literature. Immunostaining for specific matrix proteins is proposed as a direction for future work, particularly for studies aimed at clinical translation or detailed mechanistic characterization of cell-ECM interactions.

### 4.3. Alginate–dECM composite hydrogels

Beyond the decellularization, we developed hybrid hydrogels (dUECM incorporated into alginate) to serve as bioactive inks for extrusion-based 3D printing. The addition of uterine ECM markedly improved the biofunctional performance of alginate hydrogels, but it is also important to consider mechanical implications. Pure alginate hydrogels (at 3% w/v) provide reasonable mechanical strength and can be easily crosslinked with $Ca^{2+}$, but they lack cell-adhesive motifs and have a fairly inert biological profile. Conversely, pure dECM hydrogels are extremely bioactive but often quite soft and prone to contraction or rapid degradation. Our composite 3% alginate + 1.5% dUECM hydrogel aimed to balance these traits. Rheological assessments indicated that this hybrid hydrogel had favorable gelation kinetics and adequate stiffness. Encouragingly, the hybrid hydrogel's mechanical stability persists over time: after two weeks in culture conditions, the Alg–dUECM constructs retained the majority of their initial modulus. Crosslinking with $Ca^{2+}$ not only solidified the geometry but also significantly improved water retention and resistance to dissolution. TGA analysis in our study further confirmed that incorporating dUECM slightly increased the thermal stability of alginate gels, implying a denser network structure. This result echoes that blending decellularized ECM with synthetic or natural polymers can create composite hydrogels with improved mechanical integrity suitable for extrusion printing (Figure S4).[43,81,82]

It is also important to interpret our mechanical results in light of collagen integrity. The superior mechanical performance of the 3% Alg + 1.5% dUECM hydrogel (compared to alginate alone) can be mechanistically linked to the presence of collagen and other matrix proteins. Even though the ECM proteins are not crosslinked covalently in our formulation, they likely form physical entanglements and additional ionic crosslinks (via carboxyl groups binding calcium) within the alginate network.[83] Our FTIR analysis of the crosslinked hydrogel showed sharper Amide I and III bands and strong O–H stretching (indicative of extensive hydrogen bonding). The resulting Alg–dUECM formulation exhibited sufficient structural integrity to enable high-fidelity, multilayer extrusion-based 3D printing. In comparison, alginate-only hydrogels (3% w/v) displayed lower stiffness, although their printability exceeded that of dUECM-only formulations. Notably, dUECM alone could not be printed reliably, underscoring the necessity of a composite strategy. Overall, the Alg–dUECM hydrogel has promising mechanical strength, which is essential for bioengineered uterine tissue.

### 4.4. Bioactivity and cellular response in hybrid hydrogels

A primary motivation for integrating uterine dECM into alginate was to impart bioactivity that alginate lacks. Alginate is inherently non-adhesive (cells do not attach to the alginate backbone due to absence of RGD or other binding motifs).[14] In our hybrid hydrogel, the dUECM component provides a rich source of cell-interactive molecules such as collagen types I and III, laminin, fibronectin, and growth factors specific to uterine matrix.[14,84] These cues dramatically improved cell behavior in our assays. We used an immortalized human myometrial cell line (hTERT-HM) to evaluate viability and proliferation on the hydrogels. On 3% Alg + 1.5% dUECM gels, cells showed high viability and robust proliferation over 7 days, significantly outperforming cells on alginate-only or lower dECM-content gels (which had slower growth and more apoptosis). Live/Dead fluorescence imaging revealed that hTERT-HM cells adhered and spread on the hybrid hydrogel surface, whereas on plain alginate they remained round and tended to cluster with limited attachment. Moreover, the presence of uterine ECM appeared to encourage the myometrial cells to assume a spindle-like morphology and even align in parallel orientations. By day 7, cells on the Alg–dUECM hydrogel formed interwoven, multi-layered sheets reminiscent of the smooth muscle layering in native myometrium.[2] This emergent alignment is likely driven by contact guidance and mechanosensitive interactions with ECM fibrils embedded within the hydrogel, as collagenous networks can provide both topographical cues and integrin-mediated biochemical signaling that support smooth muscle–like behavior.

Immunohistochemical staining showed strong and well-organized α-SMA expression in cells cultured on Alg–dUECM hydrogels, particularly at later time points. Because α-SMA is a key marker of contractile smooth muscle cells, its sustained expression suggests maintenance of a myometrial-like phenotype. In contrast, cells on alginate-only hydrogels displayed weaker and more diffuse α-SMA staining, consistent with limited cell–matrix interactions and reduced cytoskeletal organization. The enhanced α-SMA expression in the hybrid hydrogel likely reflects combined biochemical and mechanical cues provided by the dUECM, including collagen-mediated integrin signaling and matrix stiffness within a physiologically relevant range. These features support cytoskeletal organization and reinforce the suitability of Alg–dUECM constructs for *in vitro* modeling of uterine smooth muscle tissue.

Importantly, the observed preservation of smooth muscle phenotype aligns with prior reports demonstrating the ability of hTERT myometrial cells to recellularize decellularized uterine matrices. Consistent with previous studies,[34,85] hTERT cells exhibit robust viability and phenotypic stability when cultured within uterine-derived ECM environments. However, earlier work relied on inert wool glass 3D meshes, which, despite supporting cell attachment, lack the bioactivity and degradability required for translational tissue engineering applications.[14,85] In contrast, the bioactive alginate–dUECM hydrogel not only supports cell attachment but also actively enhances proliferation and viability, offering a more functional platform for uterine tissue reconstruction.

### 4.5. Clinical and immunological considerations

For eventual clinical translation of a bioengineered uterine construct, several practical considerations must be addressed: species source of the matrix, sterilization, immunogenicity, and *in vivo* functionality. In this study, we utilized decellularized uterine tissue from a non-human source (porcine) as the ECM component. While porcine uterine ECM is biochemically similar to human (rich in collagens I/III, elastin, laminin, *etc.*), there are species-specific molecules to consider. One such factor is the galactose-α(1,3)-galactose (α-Gal) epitope present in pig tissues, which can elicit hyperacute immune reactions in primates.[86,87] Our decellularization protocol, like many others, likely removed cellular constituents carrying α-Gal, but residual α-Gal on matrix proteins could still pose a risk. Encouragingly, complete removal of cells and thorough washing (as evidenced by DNA quantification and histology) is expected to eliminate most xenoantigens. Notably, effective antigen clearance was strongly dependent on detergent treatment, as only SDS-based decellularization achieved complete histological acellularity and robust removal of xenogeneic antigens.[88,89] Although α-Gal was not directly quantified in this study, the low residual DNA content and absence of detectable nuclei within the scaffolds indicate effective removal of cellular material and associated antigens. Direct assessment of α-Gal epitopes and other residual xenoantigens is warranted in future studies to further confirm the immunological safety of the decellularized matrix.

## 5. Conclusion

The fabrication of functional uterine tissue constructs remains limited by the lack of biomaterials that are both biologically instructive and mechanically suitable for 3D printing. While dUECM offers promising bioactivity, it is not printable in its native form. Moreover, traditional multi-step decellularization methods are inefficient and can compromise ECM quality.

In this study, we developed a one-step decellularization protocol using 1% Triton™ X-100 and 1% SDS for 48 h. This approach effectively removed immunogenic cellular components while preserving key ECM molecules, eliminating the need for DNase and reducing processing time and chemical exposure. To enable 3D bioprinting, dUECM was blended with alginate, a polymer valued for its printability and mild crosslinking, to form a composite hydrogel.

Among the tested formulations, the ink containing 3% alginate with 1.5% dUECM exhibited the best performance in terms of printability, mechanical integrity, and degradation stability. *In vitro* studies using casted hydrogels confirmed that hTERT-HM cells remained viable, proliferated significantly, and displayed elongated morphology indicative of uterine smooth muscle behavior.

In conclusion, this study demonstrates that the optimized decellularization protocol can efficiently preserve the biofunctionality of uterine ECM, and that the resulting dUECM–alginate hydrogel supports key structural and cellular features necessary for uterine tissue engineering. Together, these findings lay the groundwork for future development of patient-specific, biofabricated uterine constructs for medical applications.

## Acknowledgments

The author sincerely acknowledges Dr. Ewa I. Miskiewicz for her exceptional expertise, generous training, and unwavering support in the laboratory, which has been invaluable to this research. Special thanks to BioRender.com for providing access to their platform for graphical illustrations. The author also extends deep gratitude to Dr. Amin Babaei-Ghazvini for his invaluable training on FTIR, TGA, and the tensile machine, as well as his insightful guidance on sample analysis. Additionally, heartfelt appreciation is given to the Institut Catholique des Arts et Métiers (ICAM), France, and its intern students, Alex Guinle and Marius Pannetier, for their assistance in the tissue engineering lab and significant contributions to the 3D-printing experiments.

## Funding

The authors declare that financial support was received for the research, authorship, and/or publication of this article. The support from the Natural Sciences and Engineering Research Council (NSERC) of Canada (Funding Numbers: RGPIN 06369-2019 and 2020-05315) and the University of Saskatchewan's Devolved Scholarship to the present work is acknowledged.

## Conflict of interest

Xiongbiao Chen serves as one of the Guest Editors of this special issue, but was not in any way involved in the editorial and peer-review process conducted for this paper, directly or indirectly. The authors declare they have no competing interests.

## Author contributions

*Conceptualization:* Abbas Fazel Anvari-Yazdi, Kobra Tahermanesh
*Data curation:* Abbas Fazel Anvari-Yazdi, Maryam Ejlali, Louison Blivet-Bailly, Vatsal Singh
*Formal analysis:* Abbas Fazel Anvari-Yazdi, Maryam Ejlali
*Funding acquisition:* Ildiko Badea, Xiongbiao Chen

*Investigation:* Abbas Fazel Anvari-Yazdi
*Methodology:* Abbas Fazel Anvari-Yazdi, Kobra Tahermanesh
*Project administration:* Xiongbiao Chen
*Resources:* Bishnu Acharya, Daniel J. MacPhee, Ildiko Badea, Xiongbiao Chen
*Software:* Abbas Fazel Anvari-Yazdi, Maryam Ejlali
*Supervision:* Ildiko Badea, Xiongbiao Chen
*Validation:* Abbas Fazel Anvari-Yazdi, Kobra Tahermanesh, Daniel J. MacPhee, Ildiko Badea, Xiongbiao Chen
*Visualization:* Abbas Fazel Anvari-Yazdi
*Writing–original draft:* Abbas Fazel Anvari-Yazdi
*Writing–review & editing:* Abbas Fazel Anvari-Yazdi, Kobra Tahermanesh, Maryam Ejlali, Vatsal Singh, Daniel J. MacPhee, Ildiko Badea, Xiongbiao Chen

## Ethics approval and consent to participate

Not applicable. The biological tissues used in this work were collected from biowaste from certified and registered slaughterhouse and thus, ethical approval is not required.

## Consent for publication

Not applicable.

## Availability of data

Raw data are available from the corresponding authors upon reasonable request.

## Further disclosure

The paper has been deposited in the following preprint servers: arXiv (https://doi.org/10.48550/arXiv.2506.15857) and preprints.org (https://doi.org/10.20944/preprints202506.1891.v1).